\documentclass[%
preprint,
 citeautoscript,
onecolumn,
 amsmath,amssymb,
 longbibliography,
aip,
floatfix,
]{revtex4-2}

\usepackage{amsmath}
\usepackage{amsthm}
\usepackage{amsfonts}
\usepackage{amssymb}
\usepackage{graphicx}
\usepackage{numprint}
\usepackage{mathtools}
\usepackage[colorlinks=true,linkcolor=blue,citecolor=blue,urlcolor=cyan,bookmarks,breaklinks=true]{hyperref}
\usepackage{enumitem}
\usepackage{graphicx}% Include figure files
\usepackage{dcolumn}% Align table columns on decimal point
\usepackage{bm}% bold math
\usepackage{xcolor}
\usepackage[font=small,labelfont=bf,
   justification=raggedright,
   format=plain]{caption}
\usepackage{subcaption}

\begin{document}

\title{Equilibrium States Corresponding to Targeted Hyperuniform Nonequilibrium Pair Statistics}

\begin{abstract}
The Zhang-Torquato conjecture [Phys. Rev. E 101, 032124 (2020)] states that any realizable pair correlation function $g_2({\bf r})$ or structure factor $S({\bf k})$ of a translationally invariant nonequilibrium system can be attained by an equilibrium ensemble involving only (up to) effective two-body interactions. To further test and study this conjecture, we consider two singular nonequilibrium models of recent interest that also have the exotic hyperuniformity property: a 2D ``perfect glass'' and a 3D critical absorbing-state model. We find that each nonequilibrium target can be achieved accurately by equilibrium states with effective one- and two-body potentials, lending further support to the conjecture. To characterize the structural degeneracy of such a nonequilibrium-equilibrium correspondence, we compute higher-order statistics for both models, as well as those for a hyperuniform 3D uniformly randomized lattice (URL), whose higher-order statistics can be very precisely ascertained. Interestingly, we find that the differences in the higher-order statistics between nonequilibrium and equilibrium systems with matching pair statistics, as measured by the ``hole'' probability distribution, provides measures of the degree to which a system is out of equilibrium. We show that all three systems studied possess the \textit{bounded-hole} property, and that holes near the maximum hole size in the URL are much rarer than those in the underlying simple cubic lattice. Remarkably, upon quenching, the effective potentials for all three systems possess local energy minima (i.e., inherent structures) with stronger forms of hyperuniformity compared to their target counterparts. Our methods are expected to facilitate the self-assembly of tunable hyperuniform soft-matter systems. 
\end{abstract}

\author{Haina Wang}
\affiliation{\emph{Department of Chemistry, Princeton University}, Princeton, New Jersey, 08544, USA}
\author{Salvatore Torquato}
\email[]{Email: torquato@princeton.edu}
\affiliation{Department of Chemistry, Department of Physics, Princeton Institute of Materials, and Program in Applied and Computational Mathematics, Princeton University, Princeton, New Jersey 08544, USA}
\affiliation{School of Natural Sciences, Institute for Advanced Study, 1 Einstein Drive, Princeton, NJ 08540, USA}

\maketitle
%%%END OF FOOTNOTES%%%

%%%MAIN TEXT%%%%
\section{Introduction}
Probing and characterizing structural properties of many-body systems in and out of equilibrium is a crucial task in the understanding of a large variety of physical, chemical and biological phenomena. \cite{Brenner2007,Co08,To00b,Sp20,Hu21} 
An outstanding challenge is the determination of effective interactions in many-body systems that accurately yield equilibrium states with prescribed pair statistics.
Solving such inverse problems is a powerful way to tackle the unsolved problem concerning the realizability of prescribed functional forms of pair statistics by many-body systems.\cite{Ya61,Cos04,Uc06a,To06b,Ku07,Zh20} Such investigations also enable one to probe systems with identical pair statistics but different higher-body statistics, which is expected to shed light on the well-known degeneracy problem of statistical mechanics. \cite{Ge94, Ji10a,St19}
Moreover, such effective potentials can be used to model macromolecules and solutions,\cite{Ha12b, Km16} and to design nanoparticles that self-assemble into desired structures, thereby facilitating material discovery.\cite{Re06a,To09a,Ma20a,Sh20b}

Recently, Zhang and Torquato conjectured that any realizable pair correlation function $g_2(\mathbf{r})$ or structure factor $S(\mathbf{k})$ corresponding to a translationally invariant nonequilibrium system can be attained by an equilibrium ensemble involving only one- and two-body effective interactions at positive temperatures.\cite{Zh20}
Testing the conjecture requires the precise determination of the effective interactions for a spectrum of target systems, including those with the exotic hyperuniform property. \cite{To03a,To18a} 
Disordered hyperuniform many-body systems, which can be solid and fluid states, are unusual amorphous states of matter that lie between a crystal and liquid. 
They are like perfect crystals in the way they suppress large-scale density fluctuations, and yet are like liquids or glasses in that they are statistically isotropic with no Bragg peaks; \cite{To03a,To18a} see Sec. \ref{sec:def_hu} for precise definitions.
Disordered hyperuniform states play vital roles in a variety of different contexts.
For example, disordered stealthy ground states have been discovered \cite{Uc04b, Ba08, To15} corresponding to under soft, long-ranged interactions, which are highly degenerate and have the ``bounded-hole'' property, which is a singular characteristic for a disordered system. \cite{Zh16b, Zh17a} 
Network structures derived from disordered stealthy point patterns can achieve complete photonic band gaps and have novel optical properties that were previously not thought to be possible. \cite{Fl09b,Man13a}
Disordered hyperuniform states also arise in the eigenvalues of random matrices (such as the Gaussian unitary ensemble), \cite{Dy70,To08b,La19} ground states of free fermions \cite{To08b}, as well as one-component plasmas at positive temperatures, \cite{Ha73, Ga79, Dy62a} all of which are exotic fluid states in which the particles interact Coulombically. \cite{Le22}
Other examples include including glass formation,\cite{Ma13a, Zh16a, To18a} jamming,\cite{To07,At16a,Con17,Ri21,To21c} rigidity,\cite{Zh16a,Gh18}  biology,\cite{Ji14,Ma15} localization of waves and excitations,\cite{Fl09a,Ma13b,Zh19,Sg21} antenna or laser array designs,\cite{Ch21} self-organization,\cite{He15,He17b,Ma19} fluid dynamics,\cite{Le19b,Ding18,Du22} quantum systems,\cite{Re67,Fe98,To08b,Ab17,Cr19} and pure mathematics.\cite{Sa06,Br19a,To18d,To19,Br20} 
Because disordered hyperuniform  states combine the advantages of statistical isotropy and the suppression of density fluctuations on large scales, they can be endowed with novel physical properties.\cite{Do05d,Fl09b,Man13a,Ma13a,Ji14,He15,Ma15,Zh16a,He17b,To18a,Ma19} 

While Zhang and Torquato introduced an algorithm to draw equilibrium classical particle configurations from canonical ensembles with one- and two-body interactions that correspond to targeted functional forms for $g_2({\bf r})$ or $S({\bf k})$, the algorithm does not generate explicit forms of the potentials. \cite{Zh20} Very recently, Torquato and Wang developed an inverse methodology that determines effective interactions with unprecedented accuracy. \cite{To22} Using this procedure, they  demonstrated the realizability of $g_2(\mathbf{r})$ for all $r$ and $S(\mathbf{k})$ for all $k$ for two different nonequilibrium models, including a two-dimensional (2D) nonhyperuniform random sequential addition process and a 3D hyperuniform ``cloaked'' uniformly randomized lattice (URL). \cite{To22} However, the Zhang-Torquato conjecture remains largely untested.

In this paper, we utilize this precise inverse methodology \cite{To22} to further test and study the Zhang-Torquato conjecture for unusual nonequilibrium hyperuniform systems. Hyperuniform targets are particularly challenging because at positive temperature $T$ they require a long-ranged pair interaction $v(r)$ that must be balanced by one-body potentials to stabilize the equilibrium system. \cite{To18a} 

Prior to the development of our inverse methodology,\cite{To22} predictor-corrector methods,\cite{Le85,Ly95,So96,He18} such as Iterative Boltzmann inversion (IBI) \cite{So96} and iterative hypernetted chain inversion (IHNCI),\cite{Le84, He18} were regarded to be the most accurate inverse procedures.
Both IBI and IHNCI begin with an initial discretized (binned) approximation of a trial pair potential.
The trial pair potential at each binned distance is iteratively updated to attempt to reduce the difference between the target and trial pair statistics. 
However, IBI and IHNCI cannot treat long-ranged pair interactions required for hyperuniform targets, nor do they consider one-body interactions that stabilize hyperuniform equilibrium states \cite{To22}; see Sec. \ref{sec:def_hu} for details.
These algorithms also accumulate random errors in the binned potentials due to simulation errors in the trial pair statistics, and thus do not achieve the precision required to probe realizability problems.
Moreover, because all previous methods do not optimize a pair-statistic ``distance'' functional, they are unable to detect poor agreement between the target and trial pair statistics that may arise as the simulation evolves, leading to increasingly inaccurate corresponding trial potentials, as demonstrated in Ref. \citenum{To22}.

The inverse methodology presented in Ref. \citenum{To22} improves on previous procedures in several significant ways. 
It utilizes a parameterized family of pointwise basis functions for the potential function at $T>0$, whose initial form is informed by small- and large-distance behaviors dictated by statistical-mechanical theory.
Pointwise potential functions do not suffer from the accumulation of random errors during a simulation, resulting in more accurate interactions.\cite{To22} 
Since it has recently been established \cite{Wa20} that inverse methods that target only $g_2({\bf r})$ or only $S({\bf k})$ for a limited range of $\mathbf{r}$ or $\mathbf{k}$ may generate effective potentials that are distinctly different from the unique potential dictated by Henderson's theorem, \cite{He74}
our methodology \cite{To22} minimizes an objective function that incorporates \textit{both} the target pair correlation function $g_2({\bf r})$ and structure factor $S({\bf k})$ so that both the small- and large-distance correlations are very accurately captured. 
For hyperuniform targets, our methodology is able to optimize the required long-ranged pair potential \cite{To08a} as well as the neutralizing background one-body potential;\cite{To22} see Sec. \ref{sec:meth} for details.

To assess the accuracy of inverse methodologies to target pair statistics, we introduced \cite{To22}
the following dimensionless $L_2$-norm error:
\begin{equation}
{\cal E}= \sqrt{D_{g_2}+D_{S}},
\label{L2}
\end{equation}
where $D_{g_2}$ and $D_S$ are $L_2$ functions, given by
\begin{equation}
D_{g_2}=\rho\int_{\mathbb{R}^d} [g_{2,T}(\mathbf{r})-g_{2,F}(\mathbf{r};\mathbf{a})]^2 d\mathbf{r}, 
\label{g2-norm}
\end{equation}
\begin{equation}
D_S=\frac{1}{\rho (2\pi)^d}\int_{\mathbb{R}^d} [S_{T}(\mathbf{k})-S_{F}(\mathbf{k};\mathbf{a})]^2 d\mathbf{k},
\label{S-norm}
\end{equation}
where $g_{2,F}(\mathbf{r};\mathbf{a})$ and  $S_{F}(\mathbf{k};\mathbf{a})$ represent the final
pair statistics at the end of the optimization, which depend on the vector of potential parameters $\bf a$.
We have previously shown that our method is able to treat challenging near-critical and hyperuniform targets, \cite{To22} which previous methods cannot do.
Thus, it is the only available method to determine effective interactions for nonequilibrium hyperuniform pair statistics. 
Moreover, in cases where IBI and IHNCI are able to achieve optimized potentials, e.g., for equilibrium target pair statistics without long-range interactions, our inverse methodology generally yields $L_2$-norm errors (\ref{L2}) that are an order of magnitude smaller than those via previous methods, and reaches the precision required to recover the unique potential dictated by Henderson's theorem.\cite{He74}

We study two models of recent interest from hitherto unexplored hyperuniformity classes, including a 2D perfect glass \cite{Zh16a} and a 3D critical absorbing-state model; \cite{Co08, He15} see Sec. \ref{sec:models} for detailed descriptions of the models and Sec. \ref{sec:def_hu} for the definition of hyperuniformity classes. We show that the pair statistics of both systems can be achieved by effective potentials, which lends further support to the Zhang-Torquato conjecture. 
We define a \textit{nonequilibrium-equilibrium correspondence} to be a nonequilibrium and an equilibrium system with identical number density and pair statistics. Such correspondences have important consequences, including a capacity to explore the thermodynamic and dynamic properties of the effectively equivalent equilibrium systems, such as phase behaviors, ground states and \textit{inherent structures}, i.e. local energy minima. \cite{St82a} 
Inherent structures are properties of the energy landscape that are used to estimate various thermodynamic and dynamic properties, including heat capacity, melting point and glass transition temperature. \cite{St85a,St98} 
Furthermore, structural properties of the equilibrium states, such as nearest-neighbor probability distribution functions and percolation threshold, enable one to infer these nontrivial attributes of the nonequilibrium states, which are crucial in determining mechanical and electronic properties of materials. \cite{To02a}

The aforementioned nonequilibrium-equilibrium correspondence also enables one to probe the degeneracy of structures with the same pair statistics. \cite{Ge94, Ji10a,St19}
It is known that for a  homogeneous many-body system, one- and two-body correlations are insufficient to uniquely determine the higher-body correlation functions $g_3,g_4,...$. \cite{To06b} Equilibrium and nonequilibrium systems that possess matching pair statistics must have different higher-body statistics, and such differences reflect the specific dynamics of the nonequilibrium state. \cite{Wa20}

To study the structural degeneracy of the nonequilibrium-equilibrium correspondence, we compute higher-order statistics for the perfect glass and critical absorbing-state models, as well as those for a hyperuniform 3D cloaked URL, \cite{Kl20} whose higher-order statistics can be determined with much higher precision than those of typical fluids. The effective potential for the URL has been previously determined, but the physical properties of this nonequilibrium-equilibrium pair have not yet been thoroughly studied. \cite{To22} 
We consider the readily computable ``hole'' probability density functions $E_V(r)$ and $G_V(r)$, which are related to the probability of finding holes of radius $r$ void of particles and reflect information about all $n$-particle correlation functions $g_2, g_3, g_4, \dots$; \cite{To90c} see Sec. \ref{sec:def_higher} for precise definitions. We show that each model has the ``bounded-hole'' property, \cite{Zh17a, Gh18} i.e., the maximum hole size is bounded, and differences of hole probability functions between the nonequilibrium and equilibrium systems provides a useful \textit{nonequilibrium index}. To study the behavior of the hole probability functions on approach to the maximum hole size, we introduce a precise algorithm to compute these functions for the target URL. We apply this algorithm to show that holes near the maximum hole size in the target URL are much rarer than those in the underlying simple cubic (SC) lattice.

To study the effect of quenching on the effectively equivalent equilibrium systems, we compute the inherent structures \cite{St82a} of the effective potentials for all three models. We find that all effective potentials yield inherent structures that are of stronger forms of hyperuniformity compared to those of their corresponding target systems. Our findings are expected to facilitate the self-assembly of tunable hyperuniform soft-matter systems.

We begin by providing basic definitions and background in Sec. \ref{sec:def}. In Sec. \ref{sec:models}, we describe the model nonequilibrium hyperuniform systems. Section \ref{sec:meth} provides a sketch our inverse methodology. \cite{To22} Section \ref{sec:res} presents results for the equilibrium systems corresponding to the nonequilibrium pair statistics, including the optimized potential and configurations (Sec. \ref{sec:res_opt}), higher-order statistics (Sec. \ref{sec:res_higher}) and inherent structures (Sec. \ref{sec:res_is}). We provide concluding remarks in Sec. \ref{sec:conc}. 

\section{Preliminaries and definitions}
\label{sec:def}
\subsection{Pair statistics}
\label{sec:def_pair}
We consider many-particle systems in $d$-dimensional Euclidean space $\mathbb{R}^d$ that are completely statistically characterized by the $n$-particle probability density functions $\rho_n(\mathbf{r}_1,...,\mathbf{r}_n)$ for all $n\geq 1$. \cite{Ha86} In the case of statistically homogeneous systems, $\rho_1(\mathbf{r}_1)=\rho$ and $\rho_2(\mathbf{r}_1,\mathbf{r}_2)=\rho^2 g_2(\mathbf{r})$, $\rho$ is the number
density in the thermodynamic limit, $g_2(\mathbf{r})$ is the pair correlation function, and $\mathbf{r}=\mathbf{r}_2-\mathbf{r}_1$.
If the system is also statistically isotropic, then $g_2(\mathbf{r})$ is the radial function $g_2(r)$, where $r=|\mathbf{r}|$. The ensemble-averaged structure factor $S(\mathbf{k})$ is defined as
	\begin{equation}
		S(\mathbf{k})=1+\rho \tilde{h}(\mathbf{k}),
		\label{skdef}
	\end{equation}
where $h(\mathbf{r})=g_2(\mathbf{r})-1$ is the total correlation function, and $\tilde{h}(\mathbf{k})$ is the Fourier transform of $h(\mathbf{r})$.

For a single periodic configuration containing number $N$ point particles at positions ${\bf r}_1,{\bf r}_2,\ldots,{\bf r}_N$ within a fundamental cell $F$ of a lattice $\Lambda$, the {\it scattering intensity} $\mathcal{I}(\mathbf{k})$  is defined as
	\begin{equation}
		\mathcal{I}(\mathbf{k})=\frac{\left|\sum_{i=1}^{N}e^{-i\mathbf{k}\cdot\mathbf{r}_i}\right|^2}{N}.
		\label{scattering}
	\end{equation}
For an ensemble of periodic configurations of $N$ particles within the fundamental cell $F$, the ensemble average of the scattering intensity in the infinite-volume limit is directly related to structure factor $S(\mathbf{k})$ by
	\begin{equation}
		\lim_{N,V_F\rightarrow\infty}\langle\mathcal{I}(\mathbf{k})\rangle=(2\pi)^d\rho\delta(\mathbf{k})+S(\mathbf{k}),
	\end{equation}
where $V_F$ is the volume of the fundamental cell and $\delta$ is the Dirac delta function. \cite{To18a} In simulations of many-body systems with finite $N$ under periodic boundary conditions, Eq. (\ref{scattering}) is used to compute $S(\mathbf{k})$ directly by averaging over configurations.

\subsection{Hyperuniformity}
\label{sec:def_hu}
A hyperuniform point configuration in $d$-dimensional Euclidean space $\mathbb{R}^d$ possesses a structure factor $S(\mathbf{k})$ that goes to zero as the wave number $\mathbf{k}$ vanishes, i.e., $\lim_{\left|\mathbf{k}\right|\rightarrow 0}S(\mathbf{k})=0$, which corresponds to a local number variance $\sigma^2_N(R)$ in a spherical window of radius $R$ that grows slower than $R^d$. \cite{To03a,To18a} For hyperuniform systems whose structure factor is given by a power-law in the vicinity of the origin, i.e., 
\begin{equation}
    S(\mathbf{k})\sim |\mathbf{k}|^\alpha, \quad |\mathbf{k}|\rightarrow 0.
    \label{small_k_power}
\end{equation}
The value of the exponent $\alpha>0$ determines three different ``classes'' of hyperuniformity, \cite{To03a,Za09,To18a} i.e., 
\begin{equation}
    \sigma^2(R)\sim \begin{cases}
          R^{d-1}, \quad \alpha > 1 \text{ (class I)} \\
          R^{d-1}\ln R, \quad \alpha=1 \text{ (class II)} \\
          R^{d-\alpha}, \quad 0<\alpha<1 \text{ (class III).} \\
     \end{cases}
\end{equation}
\textit{Stealthy} hyperuniform systems, which include all perfect crystals and unusual disordered states, \cite{To16b,Zh16b} are defined to be those that possess zero intensity of the structure factor for a set of wavevectors around the origin,  \cite{To16b} i.e.,
\begin{equation}
S({\bf k})=0, \qquad \mbox{for}\; 0 \le |{\bf k}| \le K,
\label{stealth}
\end{equation}
where $K>0$. They are class I hyperuniform states that can be roughly regarded to be those in which $\alpha \rightarrow \infty$. 

To achieve equilibrium hyperuniform systems at positive $T$, which are thermodynamically incompressible, one requires the following long-ranged pair interaction in the large-$r$ limit: \cite{To18a} 
\begin{equation}
    v(r)\sim 
    \begin{cases}
    r^{-(d-\alpha)}, \quad d \ne \alpha\\
    -\ln(r), \quad d = \alpha.
    \end{cases}
    \label{v_long_hu}
\end{equation}
A notable example of which is the one-component plasma (OCP), in which identical point charges interacting via the Coulomb potential in $\mathbb{R}^d$ are immersed in a rigid, uniform background of opposite charge to ensure overall charge neutrality. \cite{Dy62a}
Such long-ranged potentials can be precisely determined via our inverse methodology; see Sec. \ref{sec:meth}. 

\subsection{Higher-order statistics}
\label{sec:def_higher}
Due to the complexity of computing and storing full pointwise information of all $n$-particle correlation functions $g_3, g_4, g_5, \dots$, this work considers higher-order statistics that are readily computable, including the \textit{conditional ``hole'' probability density function} $G_V(r)$ and $g_3$ for special triangles.

Given that a spherical region $\Omega_V(r)$ of radius $r$ is empty of particles, the quantity $\rho s_1(r)G_V(r)dr$ is the probability of finding particles in a spherical shell of volume $s_1(r)dr$, where $s_1(r)$ is the surface area of a $d$-dimensional sphere of radius $r$. The function $G_V(r)$ can be expressed in terms of integrals over all the $n$-particle correlation functions $g_2, g_3, g_4, \dots$. \cite{To90c} Importantly, $G_V(r)$ is directly related to the \textit{void-exclusion probability function} or \textit{``hole'' probability function} $E_V(r)$, which gives the probability of finding a randomly located spherical region of radius $r$ empty of particles, \cite{To02a} via the relation
\begin{equation}
    G_V(r)=-\frac{E_V'(r)}{\rho s_1(r)E_V(r)}.
    \label{ev_gv}
\end{equation}

A many-body system possesses the \textit{bounded-hole property} if $E_V(r)$ has compact support, i.e., if the \textit{maximum} hole radius $r_c$ is bounded. By contrast, typical liquids possess holes of arbitrarily large size. \cite{To90b} The bounded-hole property characterizes disordered ``stealthy'' hyperuniform states \cite{Zh17a,Gh18,Mi20} and random sequential addition at saturation, \cite{To06d,Zh13a} but has been hitherto unexplored for perfect glasses and absorbing-state models.

We also study the three-body statistics for small triangles, and especially the distribution of bond angles $\theta$. Thus, we express $g_3$ in terms of $\theta$, i.e.,
\begin{equation}
    g_3(r_1,r_2,\theta)= \frac{\rho_3\left(r_1,r_2,\sqrt{r_1^2+r_2^2-2ab\cos(\theta)}\right)}{\rho^3},
    \label{g3}
\end{equation}
where $\rho_3(r_1,r_2,r_3)$ is probability density of finding three particles that form a triangle with side lengths $r_1,r_2$ and $r_3$.

\section{Nonequilibrium hyperuniform models}
\label{sec:models}
Here, we describe the three nonequilibrium target models we consider, i.e., perfect glasses, critical absorbing-state models and URL, as well as the method of generating their corresponding configurations. They are hyperuniform systems of recent interest and have important applications in photonics engineering, \cite{Yu21} packing problems, \cite{Wi21} active matter \cite{Re14} and geoscience. \cite{Wa22a}

\subsection{Perfect glasses}
Perfect glasses are exotic amorphous states of matter with positive bulk and shear moduli that banish any crystalline or quasicrystalline phases and form unique (nondegenerate) disordered states up to trivial symmetries. \cite{Zh16a, Zh17b} 
These states can be regarded as prototypical glasses since they are out of equilibrium, maximally disordered, hyperuniform, mechanically rigid with infinite bulk and shear moduli, and remarkably prohibit the formation of crystals and quasicrystals from the ground-state manifold due to configuration-space trapping. 
The pair statistics of certain perfect glasses can by realized by equilibrium canonical ensembles, \cite{Zh20} but the explicit forms of the underlying potentials have heretofore remained unknown. 

A perfect glass is created by cooling a many-body system from high to zero $T$ with a total potential energy $\Phi_N({\bf r}^N)$: \cite{Zh16a}
\begin{equation}
    \Phi(\mathbf{r}^N)=\sum_{0<|\mathbf{k}|<K}\tilde{w}(\mathbf{k})\left[S(\mathbf{k})-S_0(\mathbf{k})\right]^2,
    \label{pg_potential}
\end{equation}
where $\mathbf{r}^N$ denotes the positions of the $N$ particles in $\mathbb{R}^d$ that are subject to optimization, $K$ is the magnitude of the largest constrained wavevector, $\tilde{w}(\mathbf{k})$ is a weight function, and $S_0(\mathbf{k})$ is the desired small-$k$ behavior of the perfect-glass structure factor.  The number of independently constrained wavevectors divided by the total number of degrees of freedom, $d(N - 1)$, is a parameter $\chi$ that measures how constrained is the system. Here, we study a class II hyperuniform perfect glass with the exponent $\alpha=1$. Following Ref. \citenum{Zh16a}, we choose the parameters $N=2500$, $\tilde{w}(\mathbf{k})=\left(K/|\mathbf{k}|-1\right)^3$, $S_0(\mathbf{k})=|\mathbf{k}|/K$, $\chi=5.1$ and $K=10/a$, where $a$ is a length scale taken to be unity. These parameters are chosen so that the position of the first peak in the perfect-glass $g_2(r)$ is close to $r=1$. We generate perfect-glass configurations by finding local minima of $\Phi(\mathbf{r}^N)$ using the low-storage BFGS algorithm, \cite{Liu89} starting from Poisson initial configurations; see Ref. \citenum{Zh16a} for further details. 

\subsection{Critical absorbing states}
Random organization models reveal how chaotically-driven nonequilibrium many-body systems can self-organize. \cite{Co08,He08,He15} Such models have been applied to study the dynamic phase transition of periodically sheared particles at the onset of irreversibility. \cite{Pi05} Hexner and Levine showed that such critical absorbing states are class III disordered hyperuniform with $\alpha=0.25$ in three dimensions. \cite{He15} 

Following Ref. \citenum{He15}, we generate critical-absorbing states as follows: Starting from a Poisson initial configuration of $N$ spherical particles with unit diameter $D$, a particle is deemed ``active'' if it overlaps with another particle. In each iteration, all active particles are given a randomly oriented displacement, whose magnitude is uniformly distributed on $[0,\sigma D/2]$, where $\sigma$ is a displacement-size parameter.
The process is repeated until an absorbing state is reached where no particles overlap, resulting in a packing of packing fraction $\phi=\rho\pi D^3/6 $, or until $10^6$ iterations are performed, implying that the system is an active state. The critical state with packing fraction $\phi_c$ is determined such that half of the initial configurations with $\rho=6\phi_c/(\pi D^3)$ reach absorbing states, while the rest are active states. In this work, we choose $N=10^5$ and $\sigma=0.1$, and estimate $\phi_c$ to be $0.205\pm 0.0005$.

\subsection{Uniformly randomized lattices}
\label{sec:models_url}
Perturbed lattices serve as important models in cosmology, crystallography and probability theory. \cite{We80, Ga02} 
Uniformly randomized lattices are simple perturbed lattices that are class I hyperuniform  with $\alpha=2$. \cite{Kl20}
For 3D URL models, each lattice point in a lattice, here taken to be the SC lattice $\mathbb{Z}^3$, is displaced by a random vector that is uniformly distributed on $[-b/2, b/2)^3$, where the scalar factor $b > 0$ is the perturbation strength. By definition, the lattice constant is set to be unity. It has been shown that the structure factor for the URL point process contains Bragg peaks that coincide with the unperturbed lattice as well as a diffuse part such that $\lim_{k\rightarrow 0}S(k)\sim k^2$. \cite{Ga04,Ki18a,Kl20} Remarkably, Klatt et al. showed that the Bragg peaks in the structure factors vanish completely, or become ``cloaked'', when $b$ takes integer values. \cite{Kl20}

In this work, we study the 3D cloaked URL with $b=1$. whose effective potential has a Coulombic asymptotic form, i.e., $v(r;\mathbf{a})\sim 1/r$, as we have previously determined. \cite{To22} Due to the independence of the particles, URLs are ideal models to study statistical structural descriptors, as they are more readily computable compared to models with correlated particles. As we will show in Sec. \ref{sec:url}, $E_V(r)$ [and thus $G_V(r)$] for this model can be numerically evaluated to very high precision via our Monte-Carlo integration technique.

\section{Inverse methodology}
\label{sec:meth}
Here we sketch the methodology that we have recently introduced \cite{To22} to determine (up to) pair interactions that yield canonical ensembles that very accurately match \textit{both} target $g_2(r)$ for all $r$ and target $S(k)$ for all $k$. Importantly, for a statistically homogeneous system equilibrated under up to two-body interactions, this technique is able to extract the unique target-generating pair potential dictated by Henderson's theorem \cite{He74}. The reader is referred to Ref. \citenum{To22} for a comprehensive description of the inverse methodology.

The methodology optimizes a parameterized isotropic potential function $v(r;{\bf a})$ that can be written as a sum of $n$ smooth pointwise basis functions, i.e.,
\begin{equation} 
v(r;{\bf a})= \varepsilon \sum_{j=1}^n f_j(r/D;a_j), 
\label{basis} 
\end{equation} 
where $f_j(r/\sigma;a_j)$ is the $j$th basis function, $a_j$ is a vector of parameters (generally consisting of multiple parameters),
${\bf a}=(a_1,a_2,\ldots,a_n)$ is the ``supervector'' parameter whose components are a collection of all components of all $a_j$'s, $\varepsilon$ sets the energy scale and $D$
is a characteristic length scale, which is taken to be unity. Examples of the basis functions include the hard core as well as superexponential-, exponential-, Yukawa- and power-law-damped oscillatory functions. The components of $a_j$ are of four types: dimensionless energy scales $\varepsilon_j$, dimensionless distance scales $\sigma_j$, dimensionless phases $\theta_j$, as well as dimensionless exponents $p_j$.

The initial form of $v(r;\mathbf{a})$ is informed by the small- and large-distance behaviors of the target pair statistics $g_{2,T}(r)$ and $S_T(k)$, as dictated by statistical-mechanical theory. \cite{To18a} For hyperuniform targets, the large-$r$ behavior of the potential is determined by Eq. (\ref{v_long_hu}).
To obtain an initial form of the small- and intermediate-$r$ behavior of $v(r;\mathbf{a})$, we numerically fit the hypernetted chain (HNC) approximation \cite{Ha86} for the target pair statistics using the aforementioned forms of basis functions. The HNC approximation is given by
 \begin{equation} 
 \beta v_{\text{HNC}}(r)= h_T(r)- c_T(r)- \ln[g_{2,T}(r)],
 \label{HNC}
\end{equation}
where $h_T(r)=g_{2,T}(r)-1$ and $c_T(r)$ is the target direct correlation function, whose Fourier transform is given by the Ornstein-Zernike integral equation \cite{Or14}
\begin{equation}
{\tilde c}_T({\bf k})= \frac{{\tilde h}_T({\bf k})}{S_T({\bf k})}.
\label{OZ}
\end{equation}

Next, a nonlinear optimization procedure \cite{Liu89} is used to minimize an objective function $\Psi({\bf a})$ based on the distance between target and trial pair statistics in both direct and Fourier spaces: 
\begin{equation}
\begin{split}
        \Psi(\mathbf{a}) &=\rho\int_{\mathbb{R}^d}
 w_{g_2}({\bf r})\left(g_{2,T}({\bf r})-g_{2}({\bf r};\mathbf{a})\right)^2
d{\bf r} \\
&+ \frac{1}{\rho(2\pi)^d}\int_{\mathbb{R}^d} w_{S}({\bf k})\left(S_T({\bf k})- S(k;\mathbf{a})\right)^2 d\mathbf{k},
\end{split}
\label{Psi}
\end{equation}
where $g_{2,T}(\mathbf{r})$ and $S_T(\mathbf{k})$ are target pair statistics, $w_{g_2}({\bf r})$ and $w_{S}({\bf k})$ are weight functions, and $g_{2}({\bf r};\mathbf{a})$ and $S(\mathbf{k};\mathbf{a})$ correspond to an equilibrated $N$-particle system under $v(r;\mathbf{a})$ at a dimensionless temperature $k_BT/\varepsilon=1$, which can be obtained from Monte-Carlo (MC) (used here with $N=500$) or molecular dynamics simulations under periodic boundary conditions. 
The optimization procedure ends when $\Psi(\mathbf{a})$ is smaller than some small tolerance $\epsilon$. 
If convergence is not achieved, then a different set of basis functions is chosen and the optimization process is repeated. 
If convergence is achieved, we check whether the effective potential robustly generates the target pair statistics for systems larger than those used during the optimization process.
We performed MC simulations under the optimized potentials with $N=2500$ for 2D systems and $N=9261$ for 3D systems, and utilized the dimensionless $L_2$ functions [(\ref{g2-norm}) and (\ref{S-norm})] and the dimensionless total $L_2$-norm error (\ref{L2}) between target and trial pair statistics to assess how close these quantities match. 

For hyperuniform targets, $v(r;\mathbf{a})$ has the long-ranged asymptotic form given by Eq. (\ref{v_long_hu}), which can be regarded as a generalized Coulombic interaction of ``like-charged'' particles. \cite{To18a} Thus, one requires a neutralizing background one-body potential to maintain stability. \cite{To18a, Ha73, Ga79, Dy62a} \footnote{Such background terms have been employed to study numerically the one-component plasma \cite{Ha73, Ga79} and the Dyson log gas. \cite{Dy62a}} Importantly, to perform the MC simulations, the total potential energy involving the long-ranged one- and two-body potentials is efficiently evaluated using the Ewald summation technique. \cite{Ew21} It is noteworthy that the optimized potentials via our methodology generally yields pair statistics that accurately match their corresponding targets with total $L_2$-norm errors that are an order of magnitude smaller than that of previous methods, in cases where previous methods are applicable. \cite{To22}

\section{Results for the nonequilibrium-equilibrium correspondence}
\label{sec:res}
\subsection{Optimized equilibrium states}
\label{sec:res_opt}
\begin{figure}[htp]
  \centering
  \subfloat[]{\label{pgAlpha1_snapTarget}\includegraphics[width=38mm]{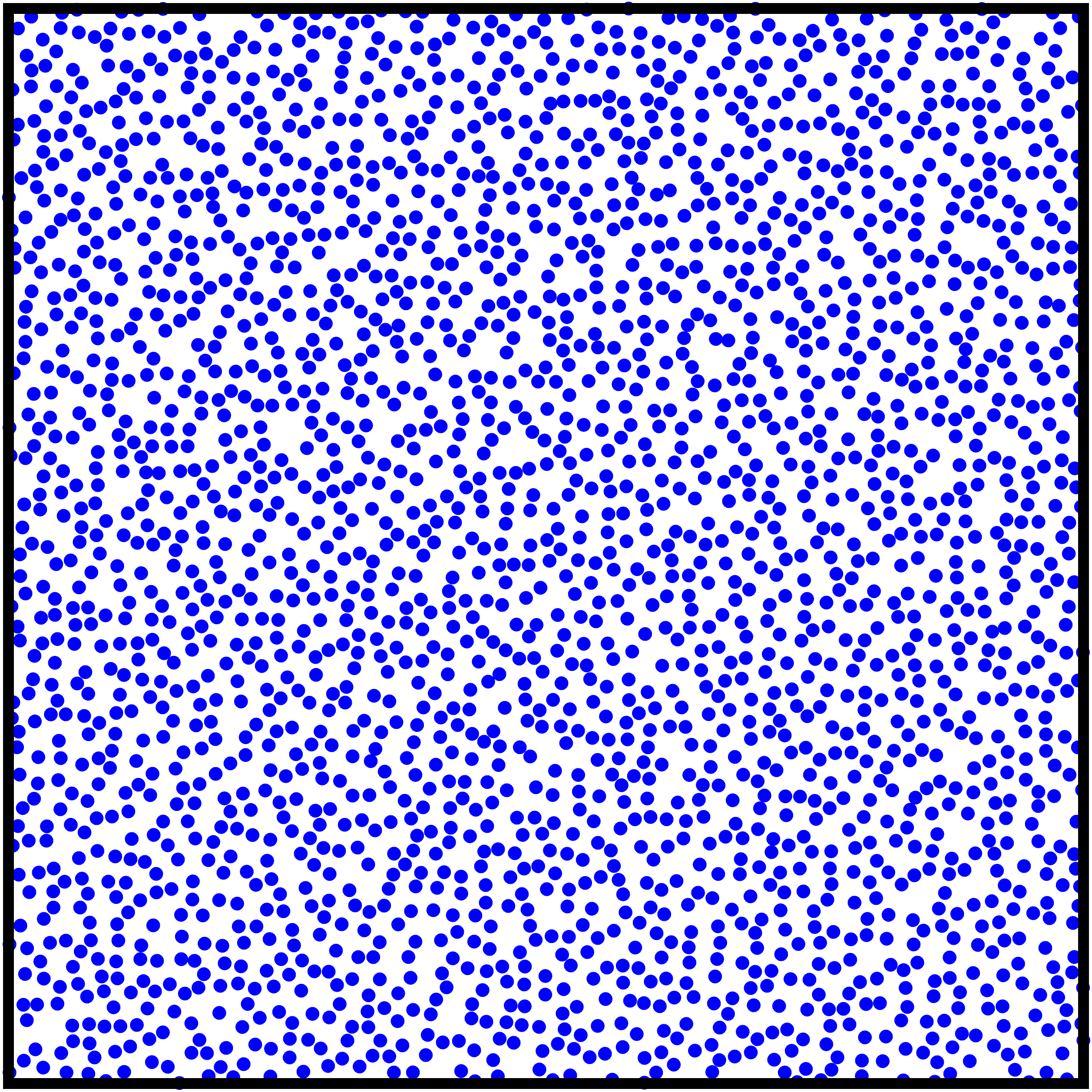}}
  \hspace{2mm}
  \subfloat[]{\label{pgAlpha1_snapInferred}\includegraphics[width=38mm]{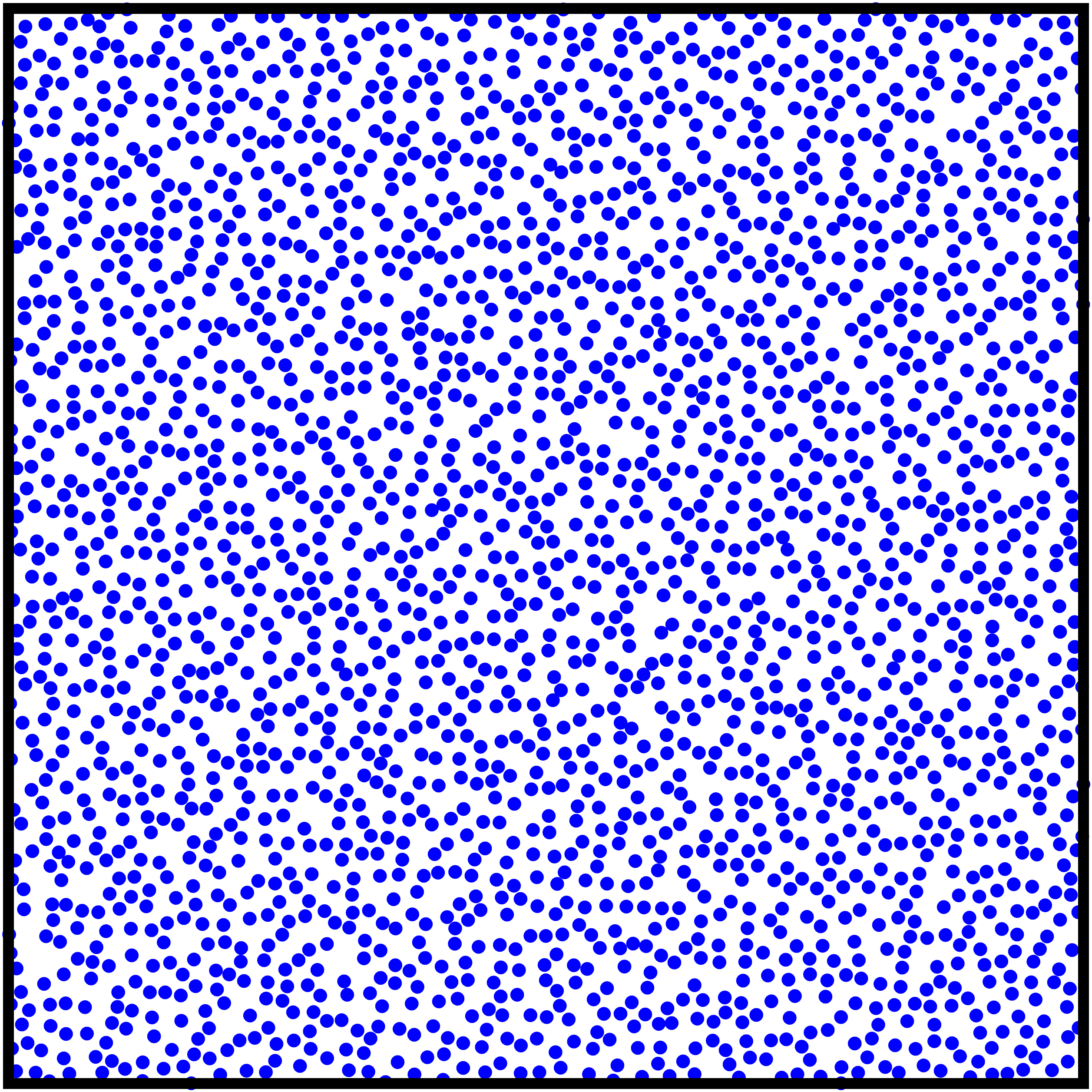}}
  
  \subfloat[]{\label{pgAlpha1_v}\includegraphics[width=40mm]{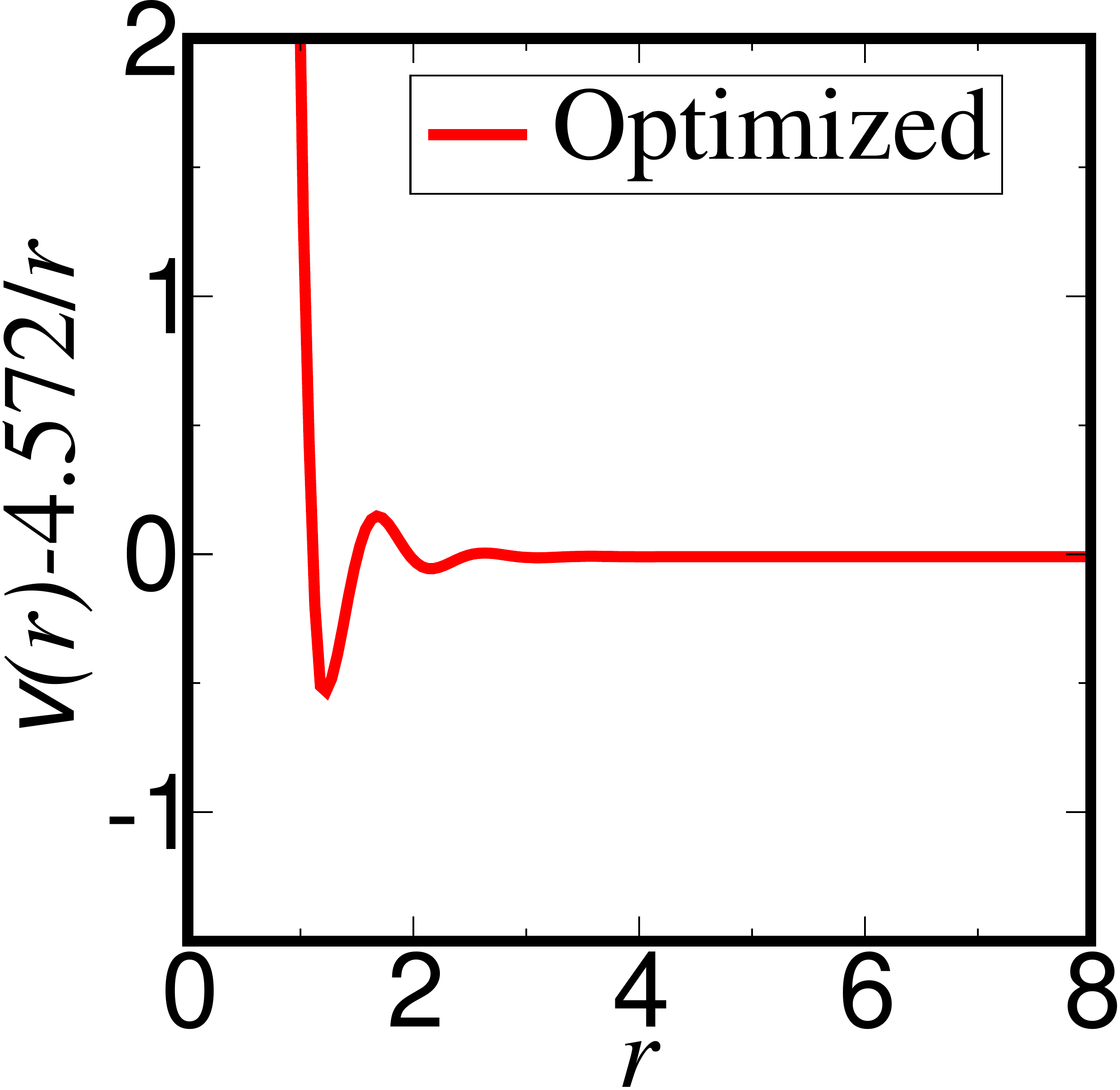}}
  \hspace{1mm}
  \subfloat[]{\label{pgAlpha1_g2}\includegraphics[width=40mm]{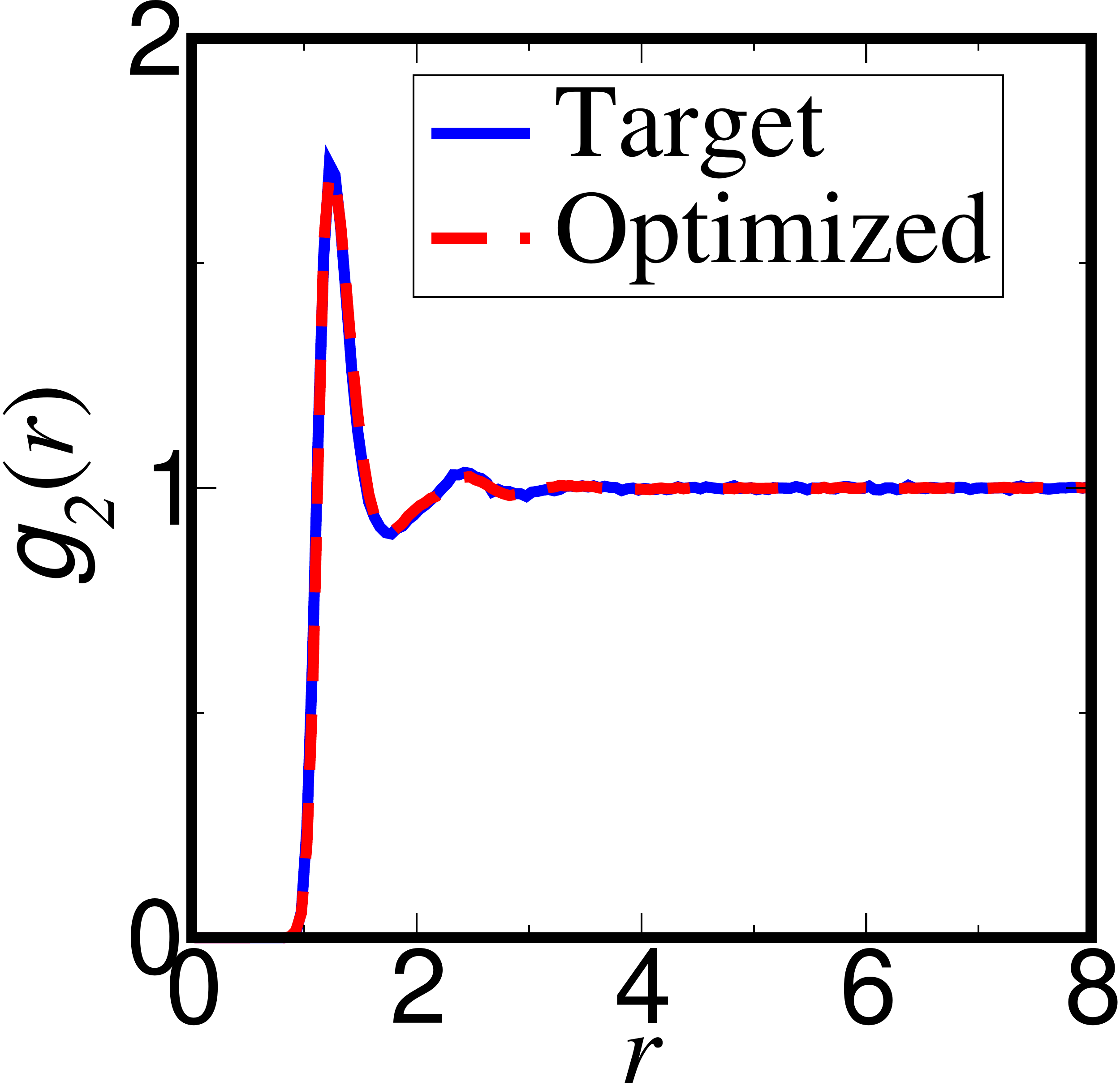}}

  \subfloat[]{\label{pgAlpha1_S}\includegraphics[width=40mm]{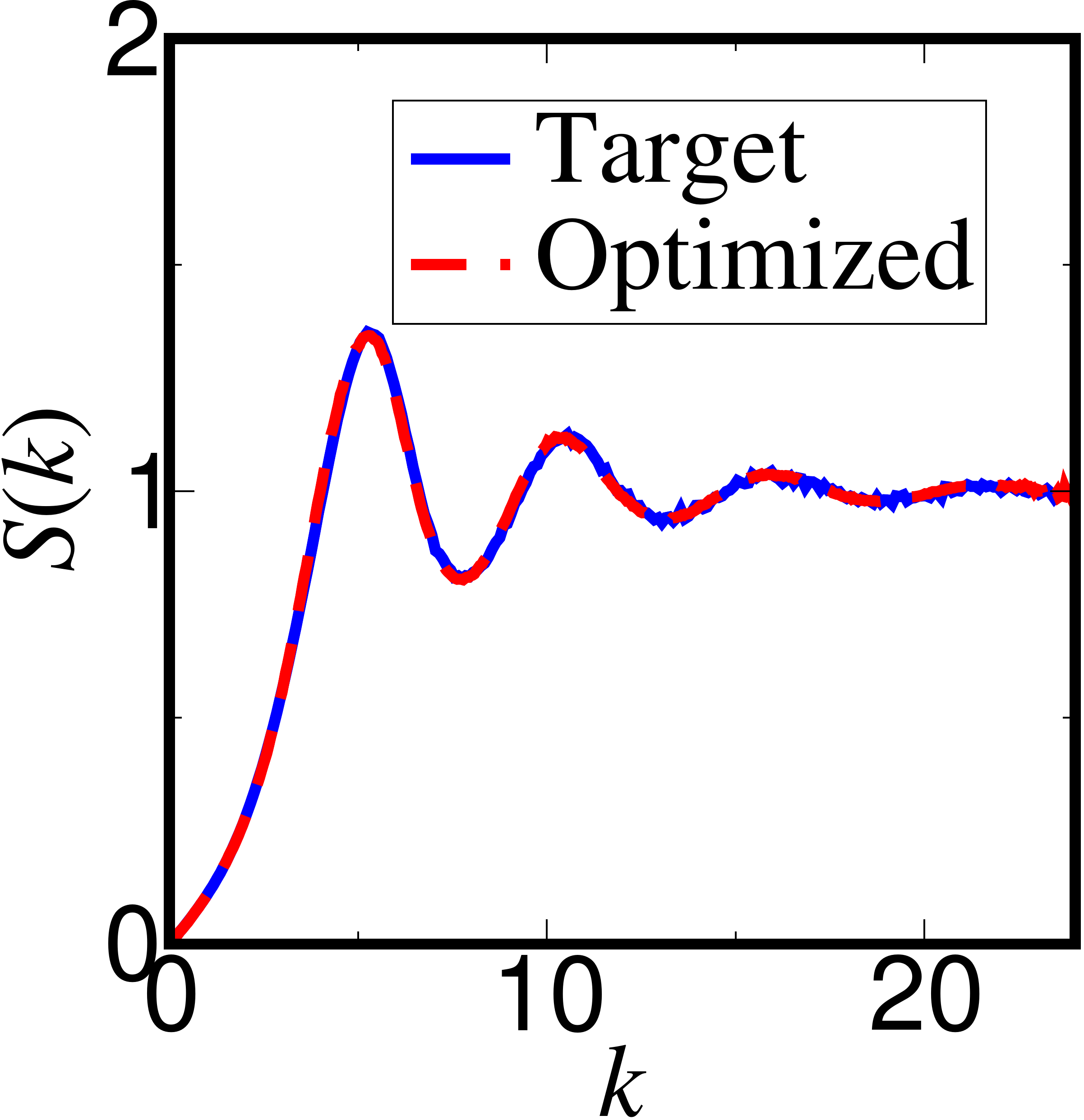}}
  \hspace{1mm}
  \subfloat[]{\label{pgAlpha1_Slog}\includegraphics[width=45mm]{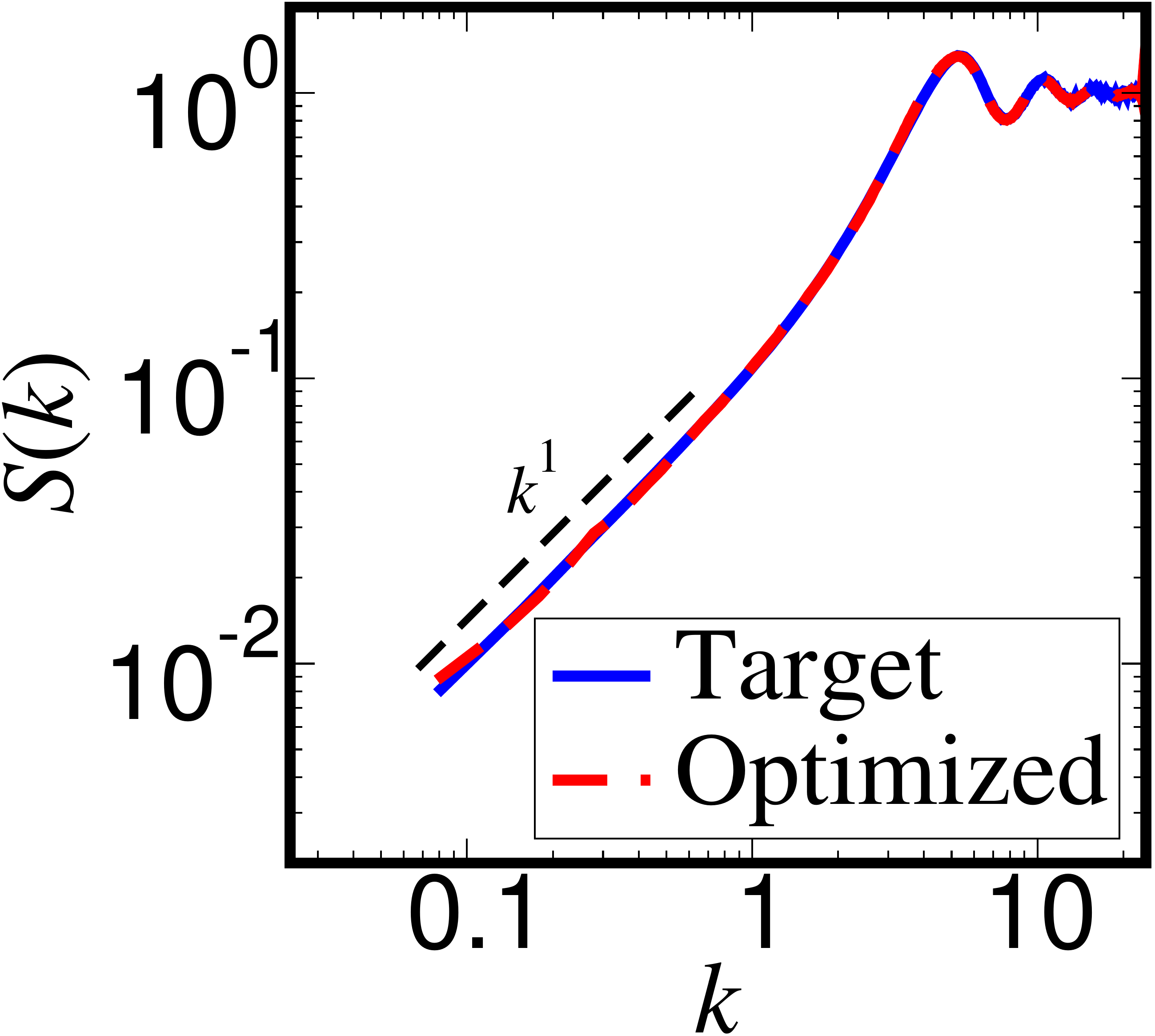}}
  \caption{(a) A 2,500-particle configuration of the 2D perfect glass. 
  (b) A 2,500-particle configuration of the optimized equilibrium state corresponding to the perfect glass. 
  (c) Short-ranged part of the optimized pair potential. (d) Targeted and optimized pair correlation functions with $N=2500$. Here we find that the $L_2$ norm function is $D_{g_2}=0.0030$.
  (e) Targeted and optimized structure factors with $N=2500$.  Here we find that the $L_2$ norm function is $D_{S}=0.0030$. The $L_2$-norm error is $\mathcal{E}=0.077$. 
  (f) Log-log plot of the targeted and optimized structure factors, showing their $k^1$ scaling behavior at small $k$.}
  \label{fig:pgAlpha1}
\end{figure}
Here, we presents the effective potentials and the pair statistics for the equilibrium states corresponding to the target nonequilibrium states. Appendix B gives the explicit functional forms of the optimized potentials.

Figure \ref{fig:pgAlpha1} shows the configurations of the target and optimized systems for the perfect glass as well as the corresponding effective pair potential and pair statistics.
As shown in Fig. \ref{fig:pgAlpha1}(a) and (b), the nonequilibrium and equilibrium configurations are visually very similar. 
The optimized effective pair potential has the expected asymptotic behavior $v(r)\sim 1/r$, as dictated by Eq. (\ref{v_long_hu}).
The short-ranged part of the potential [Fig. \ref{fig:pgAlpha1}(c)], i.e., $v(r)-4.572/r$, contains two local minima at pair distances 1.20 and 2.10, respectively. 
Figures \ref{fig:pgAlpha1}(d) and (e) show $g_2(r)$ and $S(k)$, respectively, for the target and optimized systems with $N=2500$. 
Figure \ref{fig:pgAlpha1}(f) depicts the structure factors on a log-log scale, showing that they are linear in $k$ at small $k$.
We find that the $L_2$ functions [Eqs. (\ref{g2-norm})--(\ref{S-norm})] are $D_{g_2}=0.0030$, $D_S=0.0030$ and the $L_2$-norm error (\ref{L2}) is $\mathcal{E}=0.077$, showing that the pair statistics of the perfect glass are in excellent agreement with those of the optimized equilibrium system in both direct and Fourier spaces.
Note that these errors are an order of magnitude smaller than the errors obtained via IHNCI for typical equilibrium dense liquids. \cite{To22} 
Remarkably, these results imply that one can reproduce structures that arise from two-, three and four-body interactions via effective pair interactions.

%Importantly, the equilibrium system that we discover mimics the perfect glass derived from two-, three and four-body interactions, but at the pair level.
%However, the former has nonzero entropy and thus does not preserve the unique symmetry of the perfect glass, enabling one to study its thermodynamic properties such as entropy, free energy and phase behaviors.

\begin{figure}[htp]
  \centering
  \subfloat[]{\label{randOrg3D_snapTarget}\includegraphics[width=40mm]{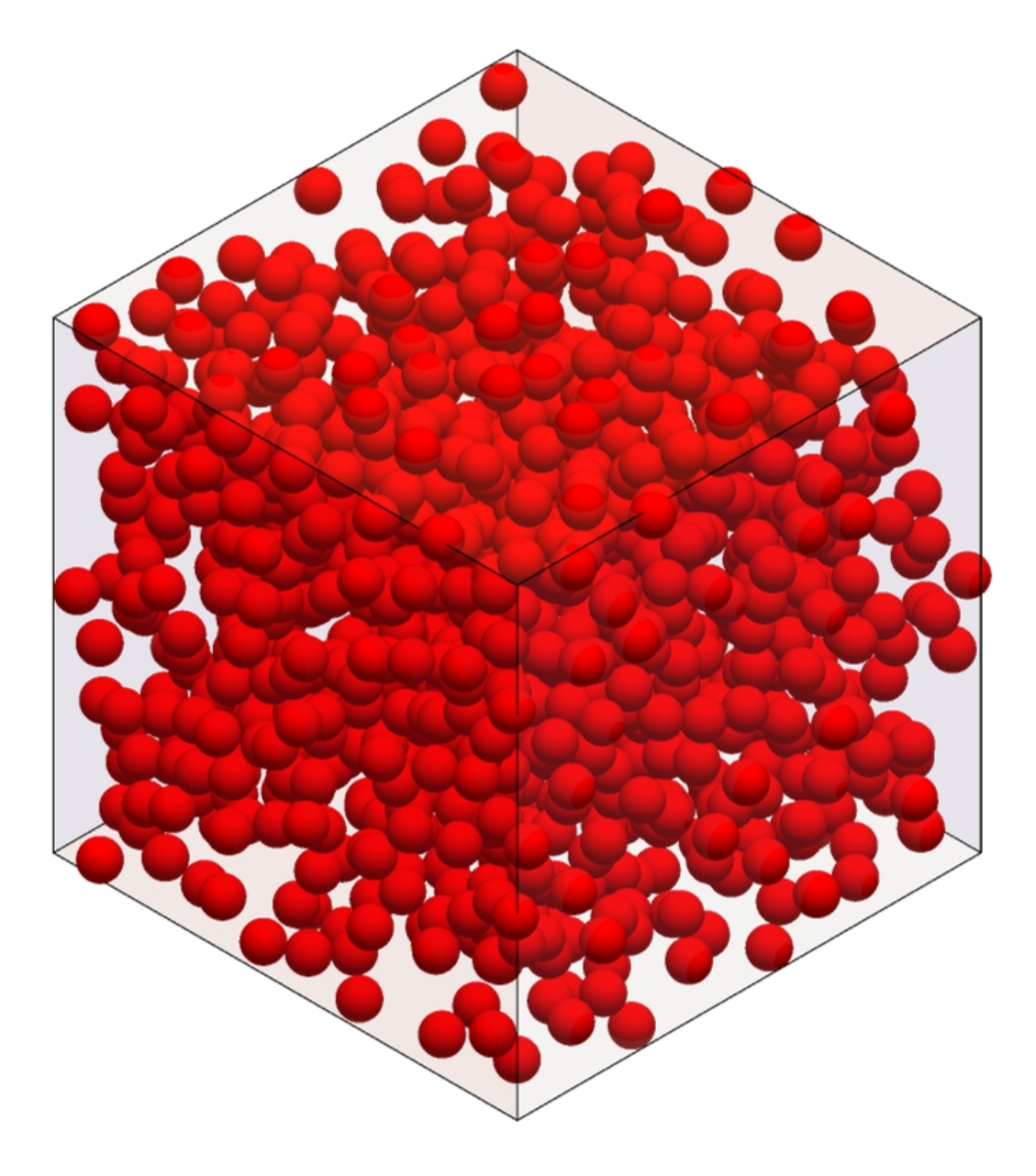}}
  \subfloat[]{\label{randOrg3D_snapInferred}\includegraphics[width=38mm]{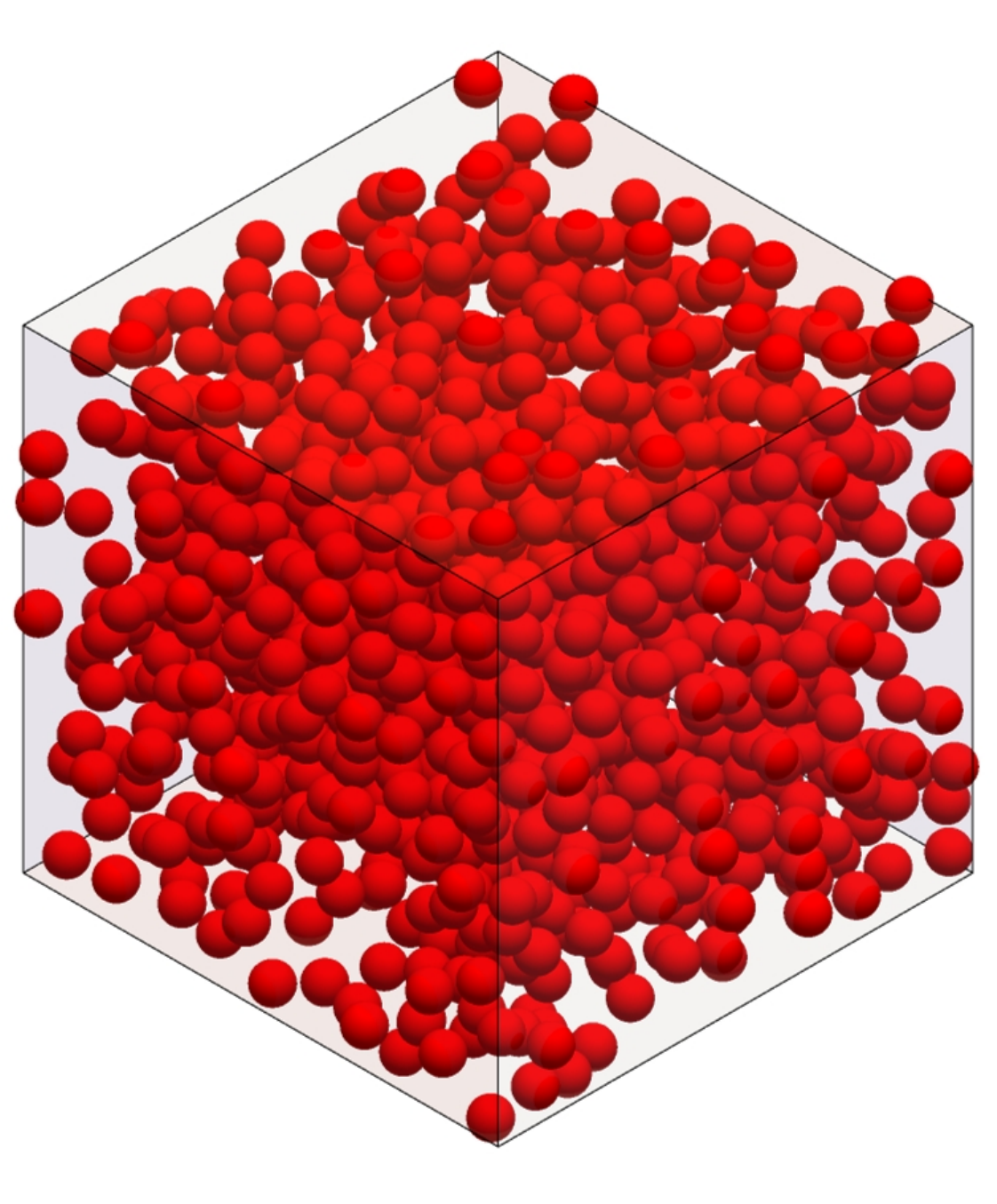}}
  
  \subfloat[]{\label{randOrg3D_v}\includegraphics[width=40mm]{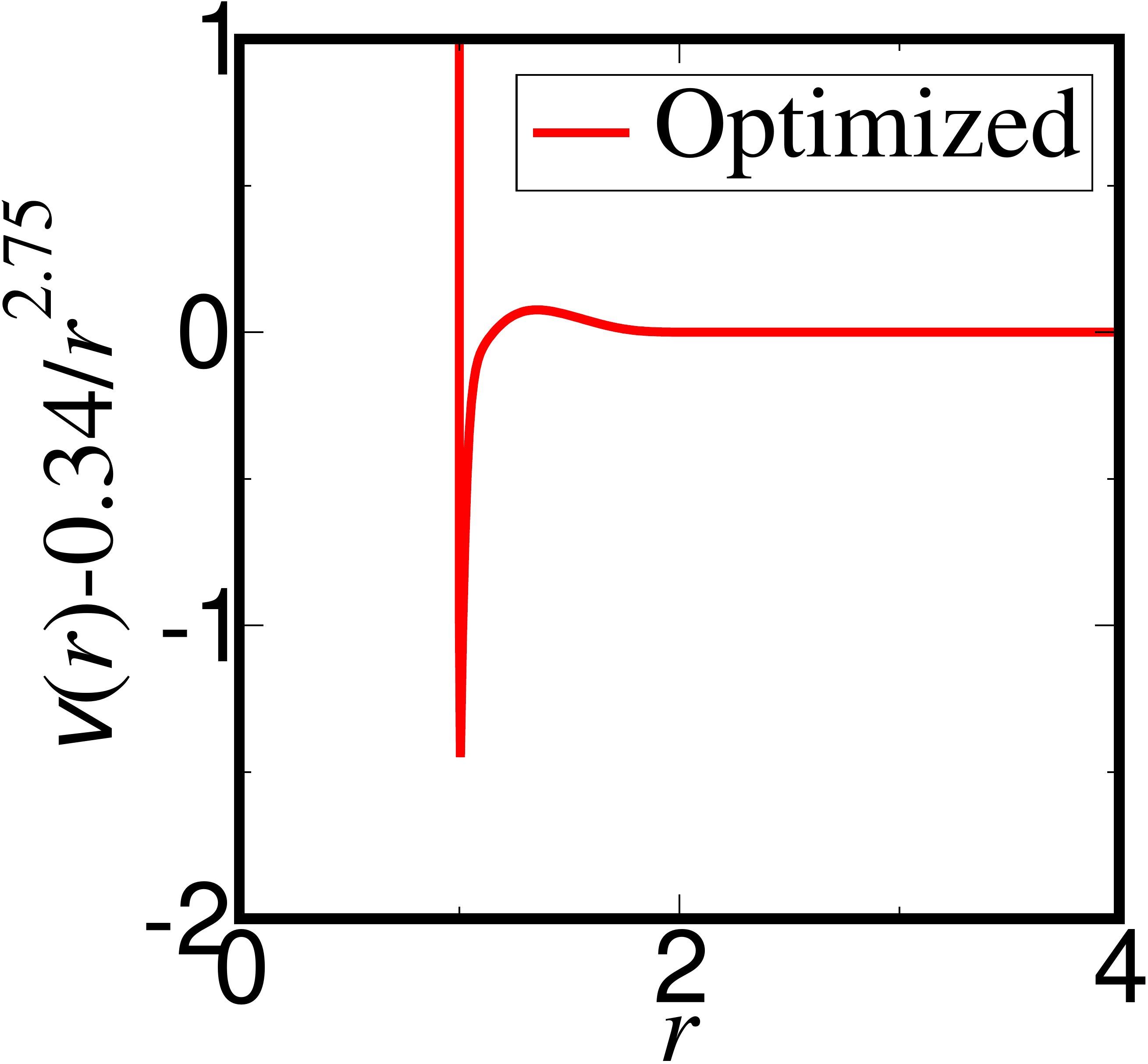}}
  \hspace{1mm}
  \subfloat[]{\label{randOrg3D_g2}\includegraphics[width=40mm]{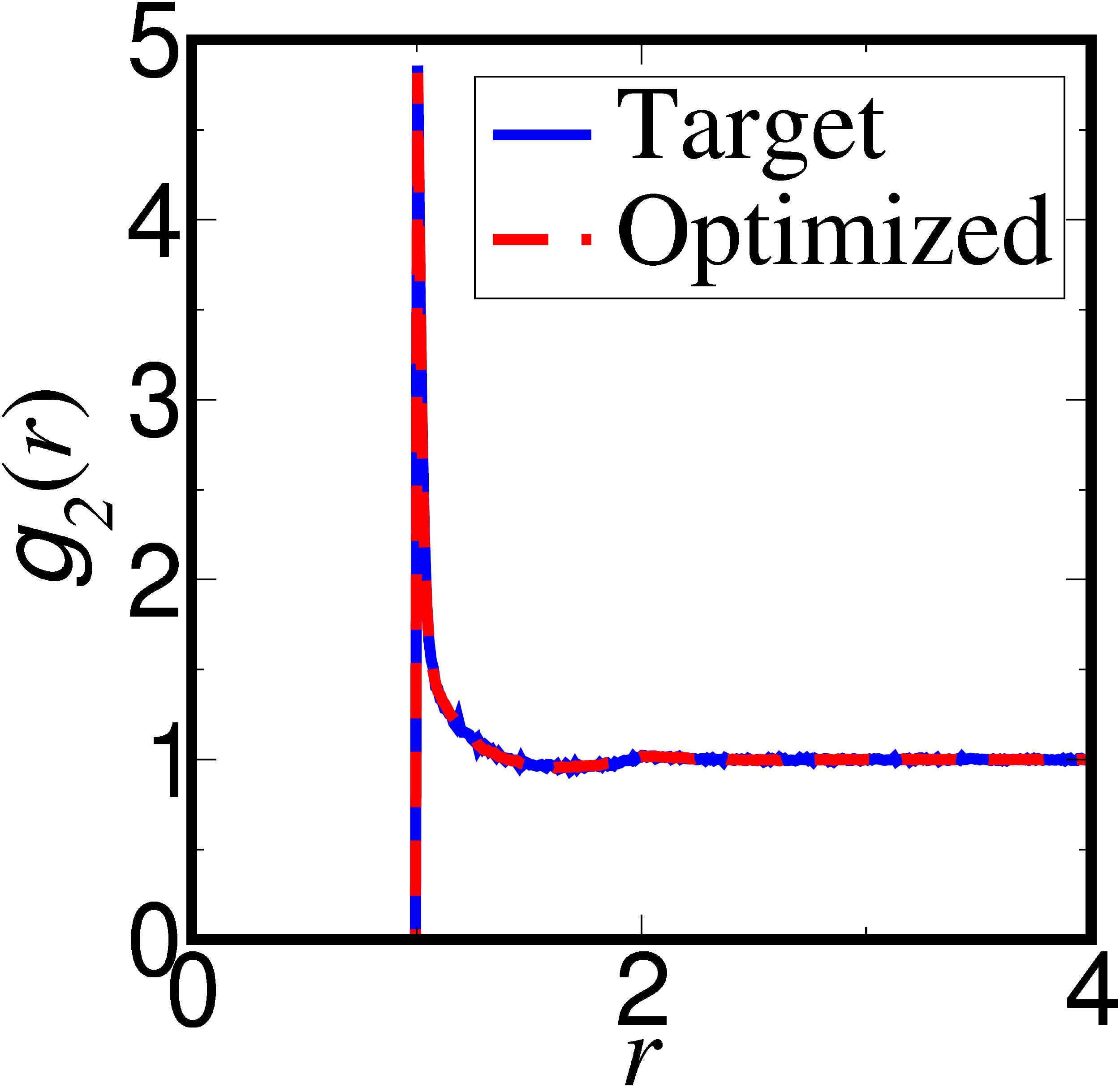}}
  
  \subfloat[]{\label{randOrg3D_S}\includegraphics[width=38mm]{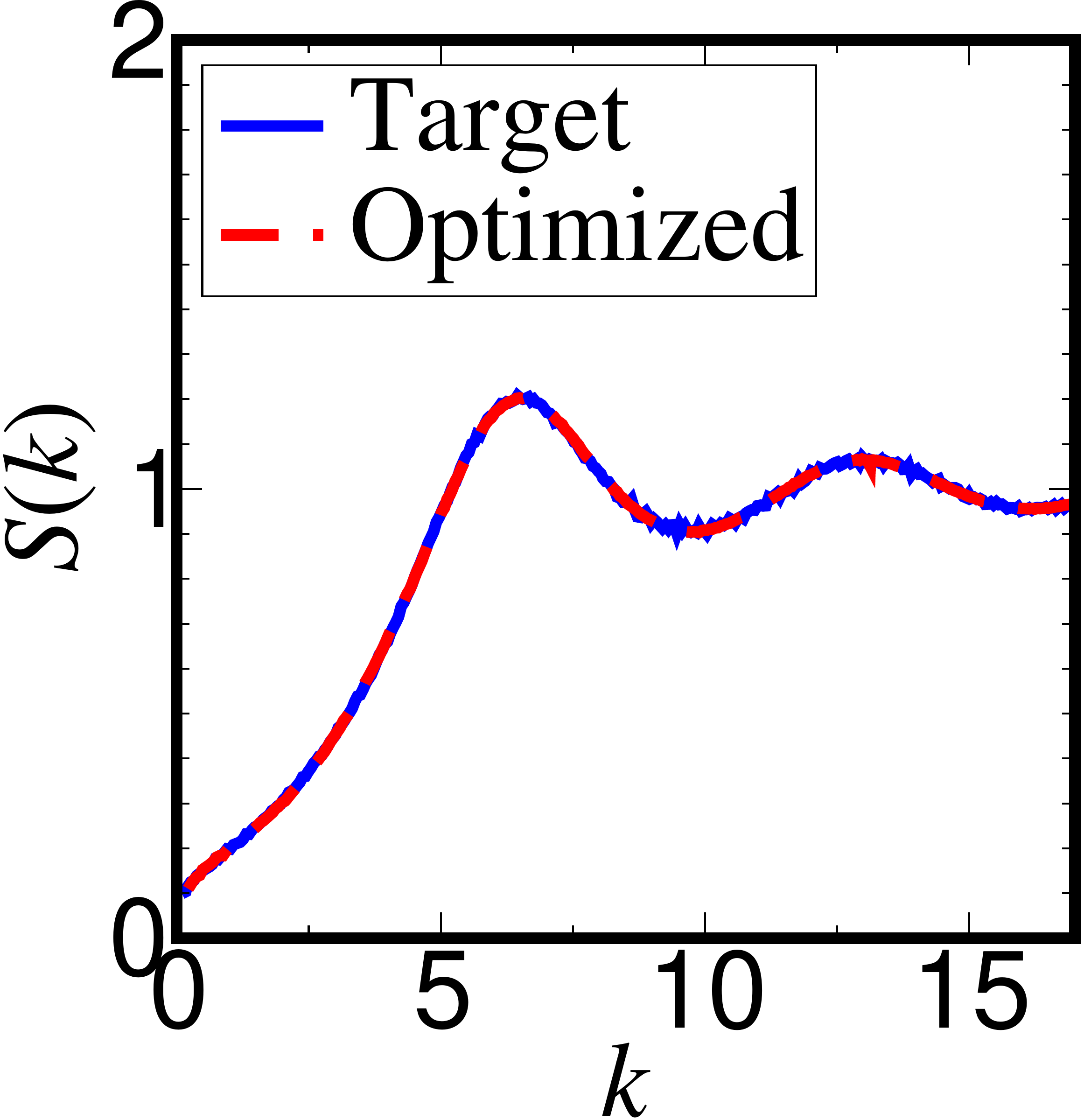}}
  \hspace{1mm}
  \subfloat[]{\label{randOrg3D_Slog}\includegraphics[width=40mm]{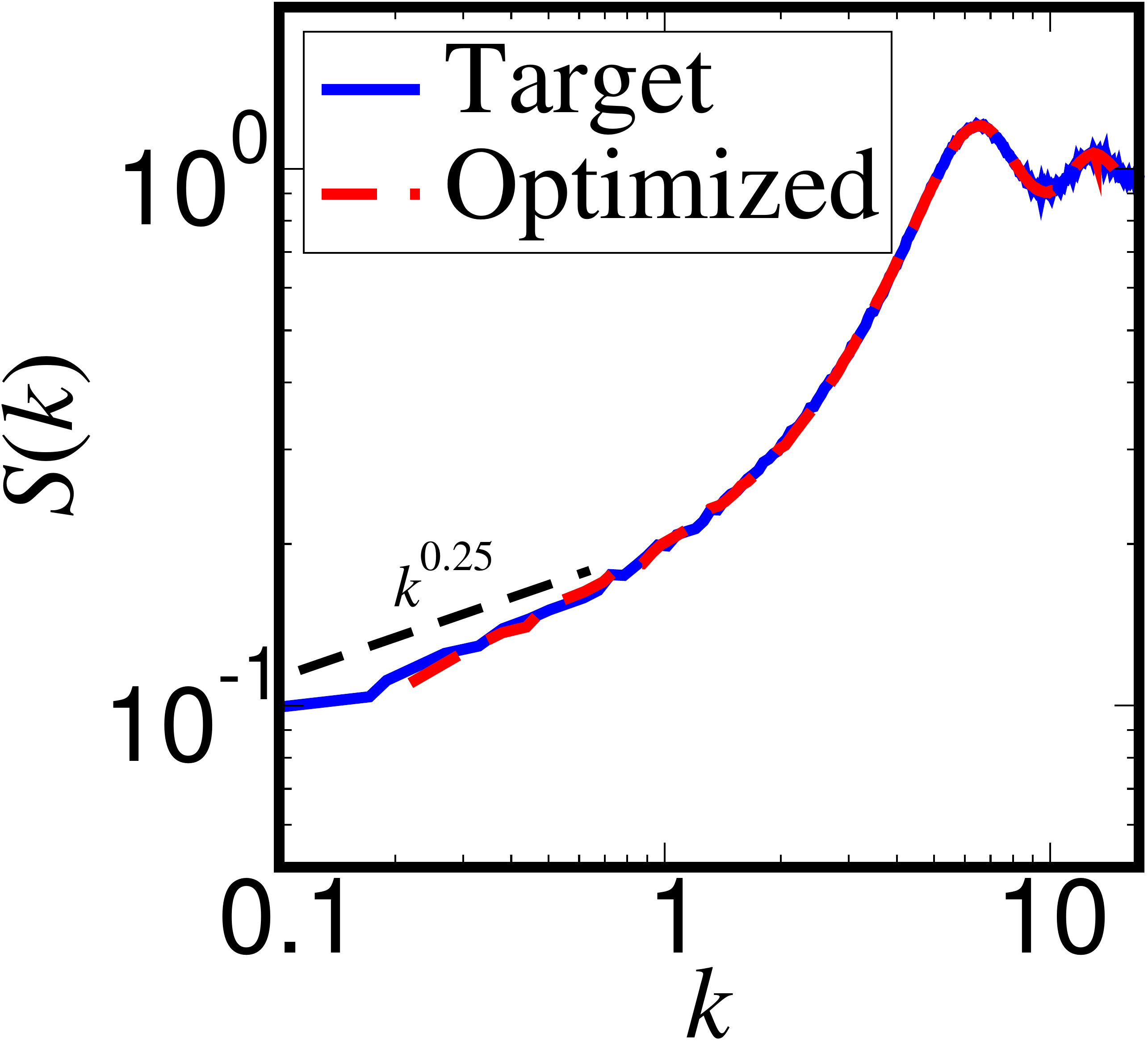}}
  \caption{(a) A 1,000-particle configuration of the 3D critical absorbing-state model. (b) A 1,000-particle configuration of the optimized equilibrium state corresponding to the critical-absorbing state. (c) Short-ranged part of the optimized pair potential. 
  (d) Targeted and optimized pair correlation functions with $N=9261$. Here we find that the $L_2$ function is $D_{g_2}=0.0020$. (e) Targeted and optimized structure factors with $N=9261$. Here we find that the $L_2$ function is $D_S=0.0022$. The $L_2$ norm error is $\mathcal{E}=0.066$. (f) Log-log plot of the targeted and optimized structure factors, showing their $k^{0.25}$ scaling behavior at small $k$.}
  \label{fig:randOrg}
\end{figure}

Figure \ref{fig:randOrg} presents the configurations, effective pair potential and pair statistics of the target and optimized systems for the absorbing-state model.
The target and optimized configurations closely resemble each other [Fig. \ref{fig:randOrg}(a) and (b)]. The effective pair potential [Fig. \ref{fig:randOrg}(c)] contains a hard core and a sharp minimum at the sphere diameter $r=1$. It decays asymptotically as $v(r)\sim r^{-2.75}$, which yields the correct the $k^{0.25}$ behavior of $S(k)$ at small $k$. 
We see that both $g_{2}(r; \mathbf{a})$ and $S(k;\mathbf{a})$ with $N=9261$ are in excellent agreement with those of the target system (Fig. \ref{fig:randOrg}(d)--(f)), as manifested by the small values of the $L_2$ functions [Eqs. (\ref{g2-norm})--(\ref{S-norm})] and the $L_2$-norm error (\ref{L2}), given by $D_{g_2}=0.0020$, $D_S=0.0022$ and $\mathcal{E}=0.066$.

\begin{figure}[htp]
  \centering
  \subfloat[]{\label{3DURL_snapTarget}\includegraphics[width=40mm]{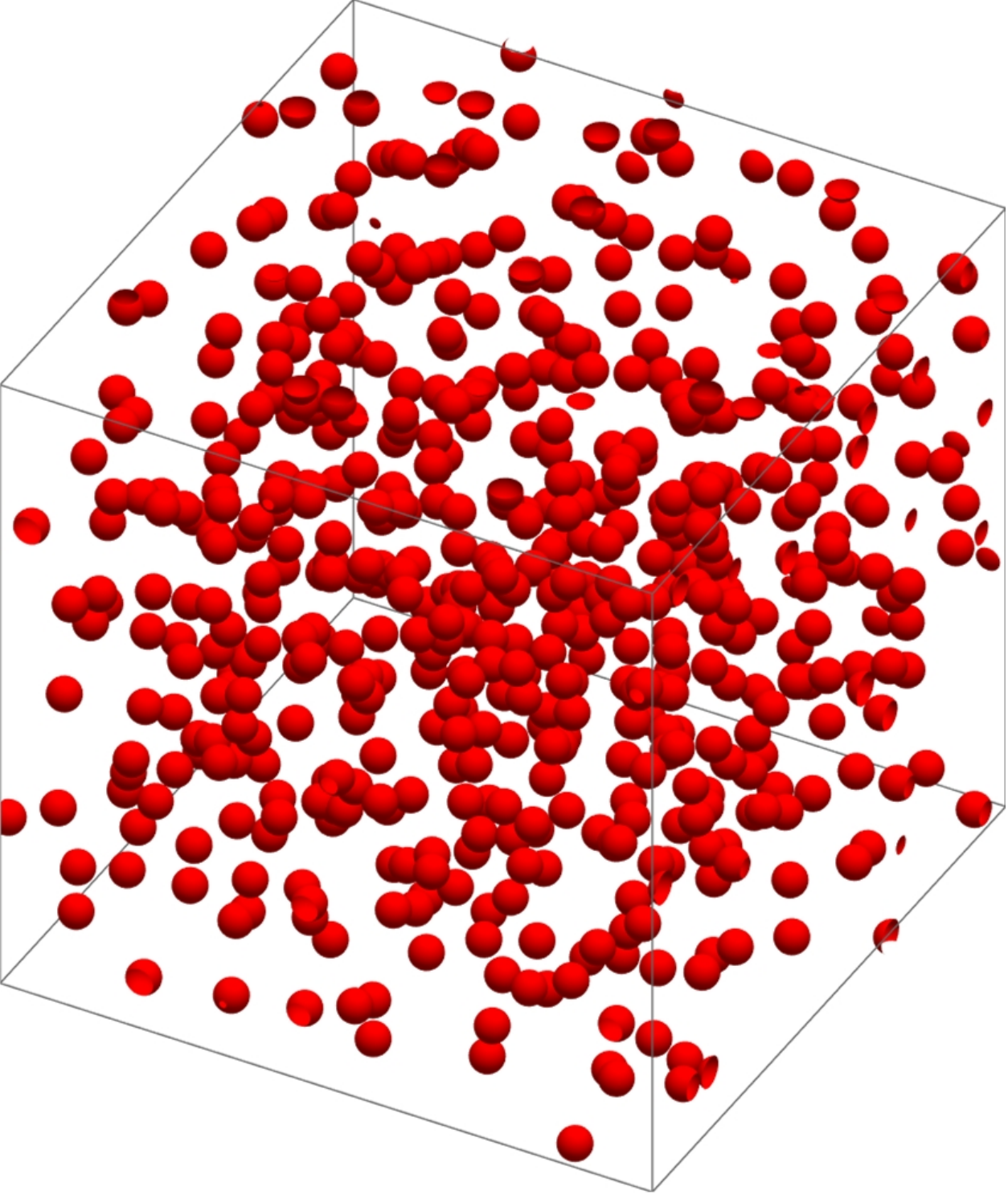}}
  \hspace{0.5em}
  \subfloat[]{\label{3DURL_snapInferred}\includegraphics[width=40mm]{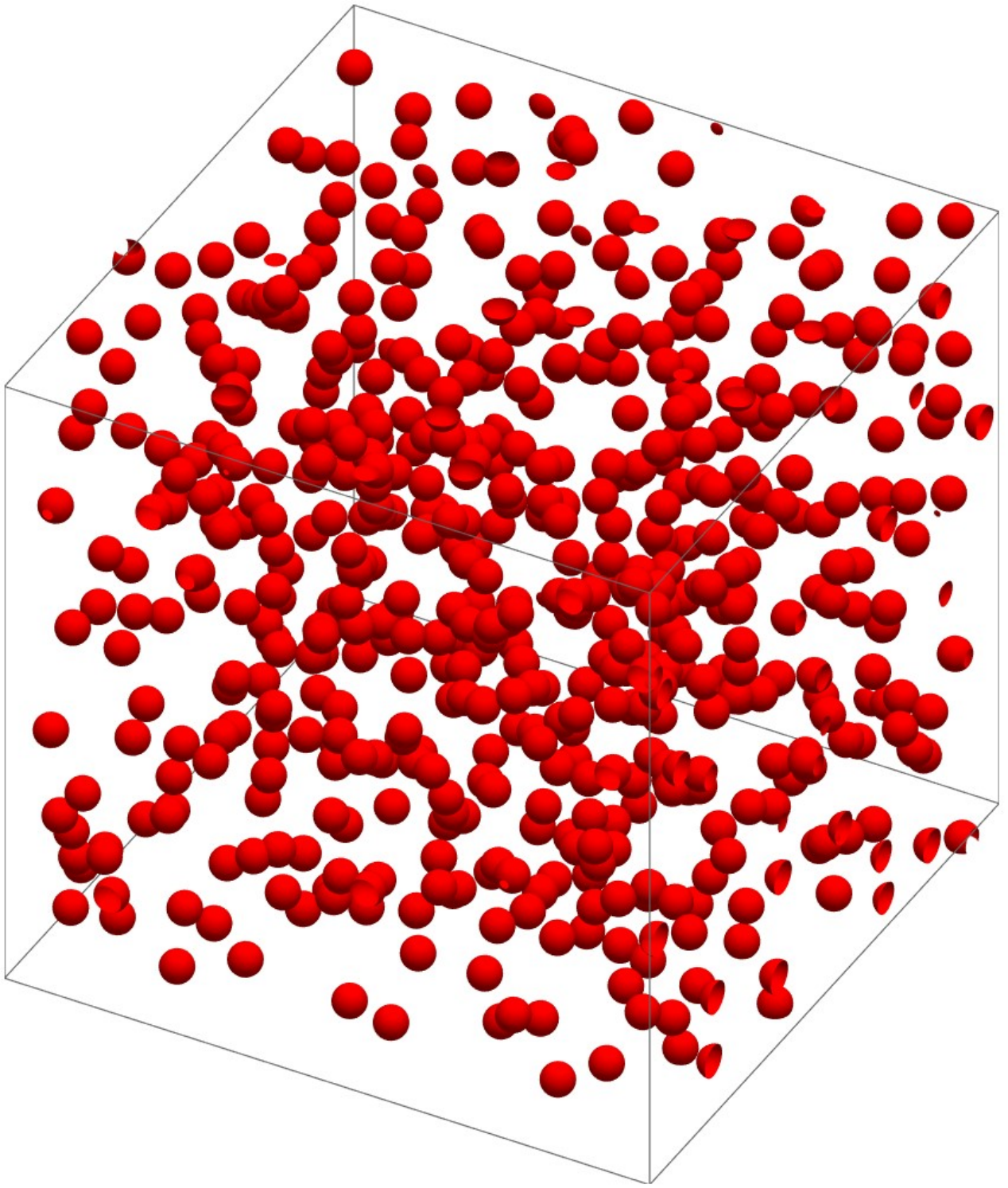}}
  
  \subfloat[]{\label{3DURL_v}\includegraphics[width=40mm]{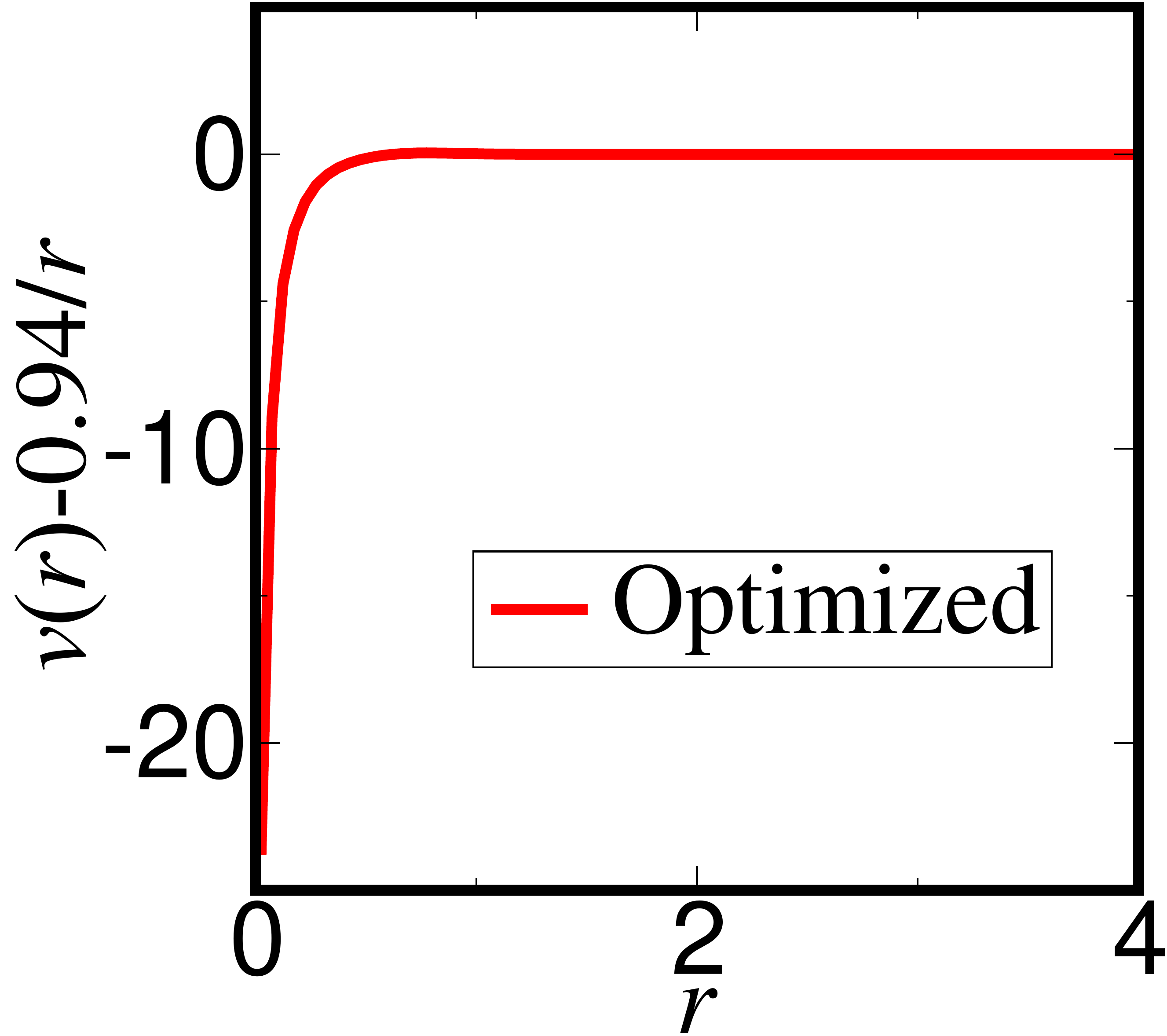}}
  \hspace{1em}
  \subfloat[]{\label{3DURL_g2}\includegraphics[width=40mm]{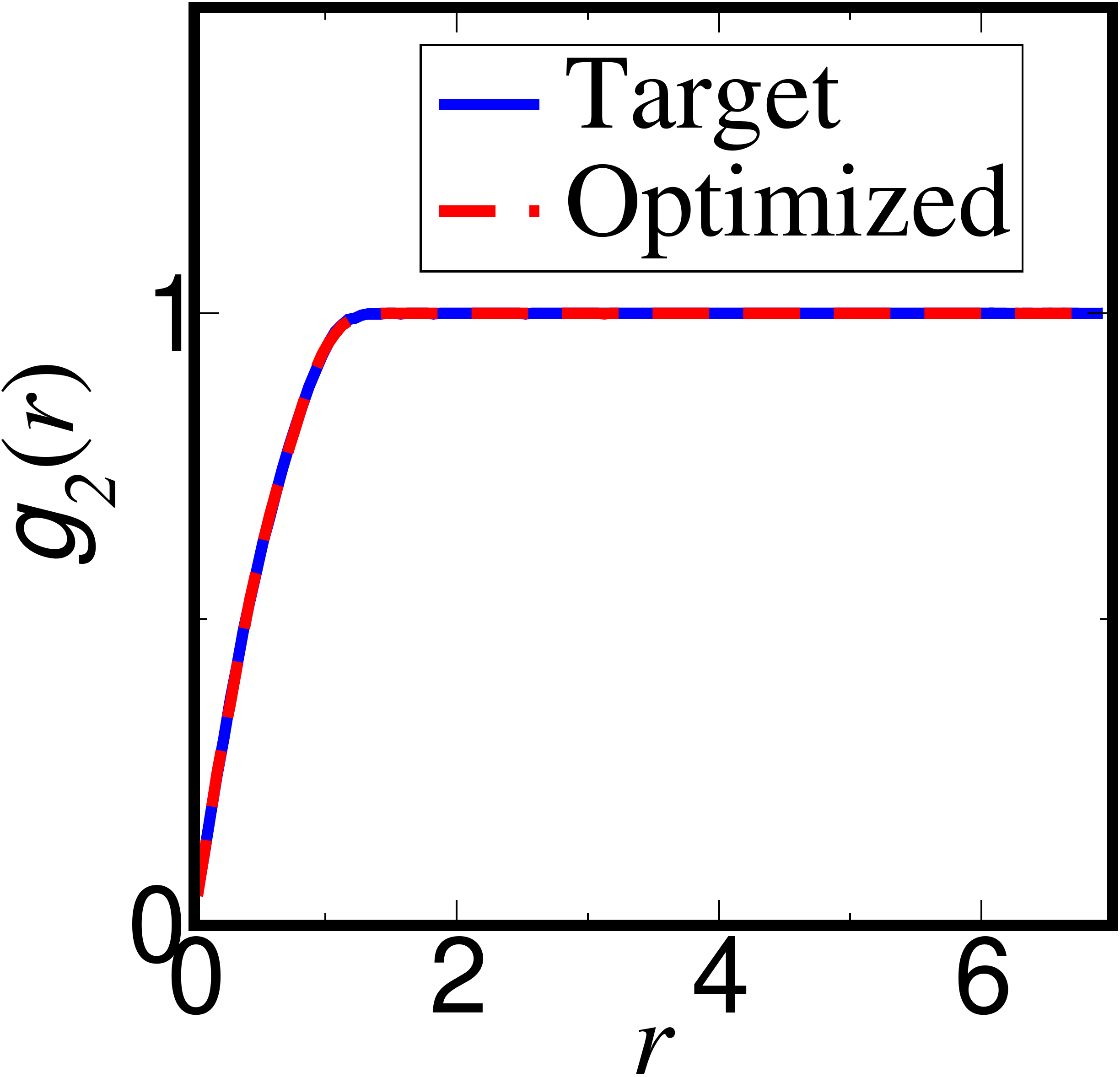}}

  \subfloat[]{\label{3DURL_s}\includegraphics[width=38mm]{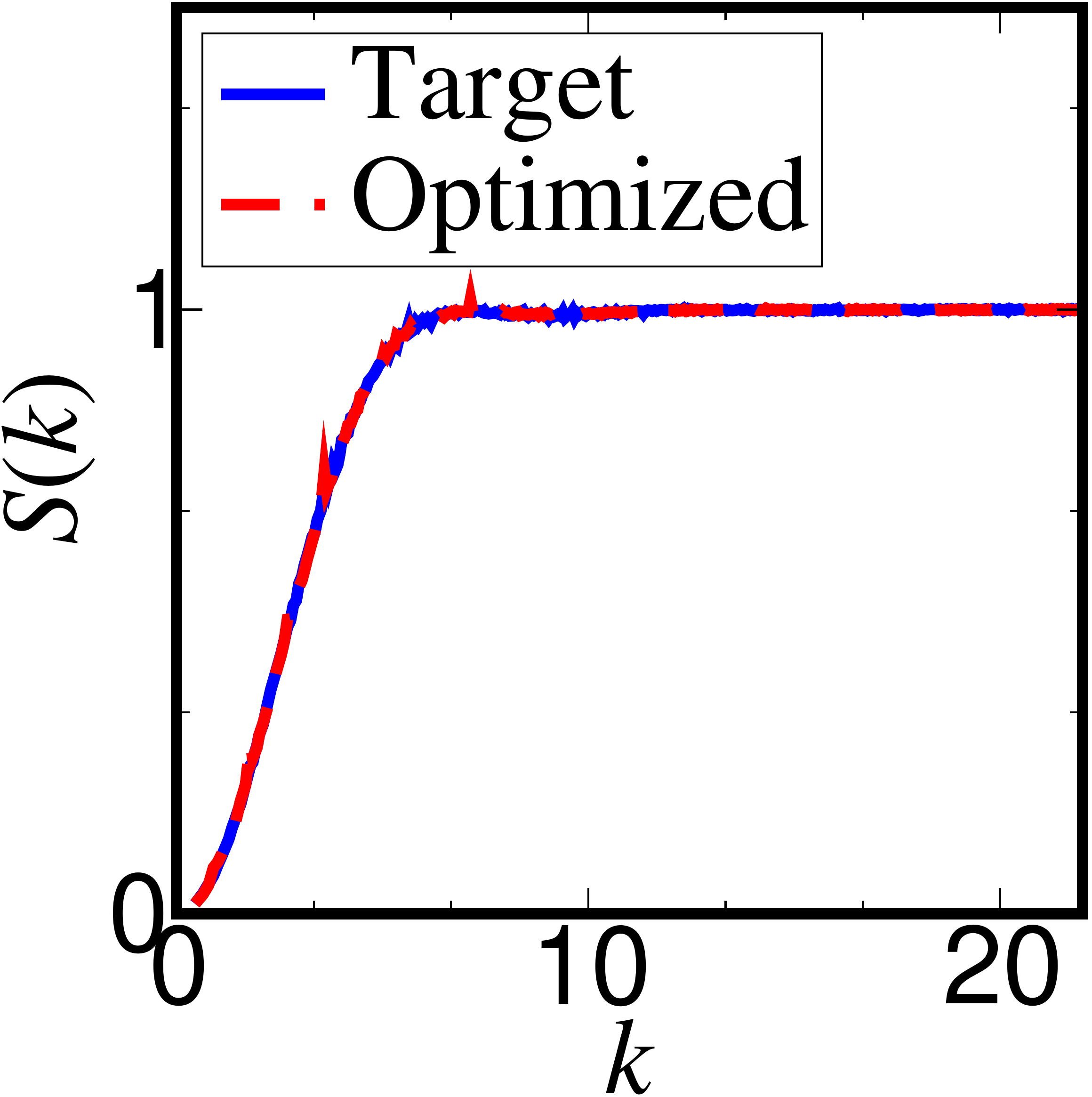}}
  \hspace{0.5em}
  \subfloat[]{\label{3DURL_slog}\includegraphics[width=42mm]{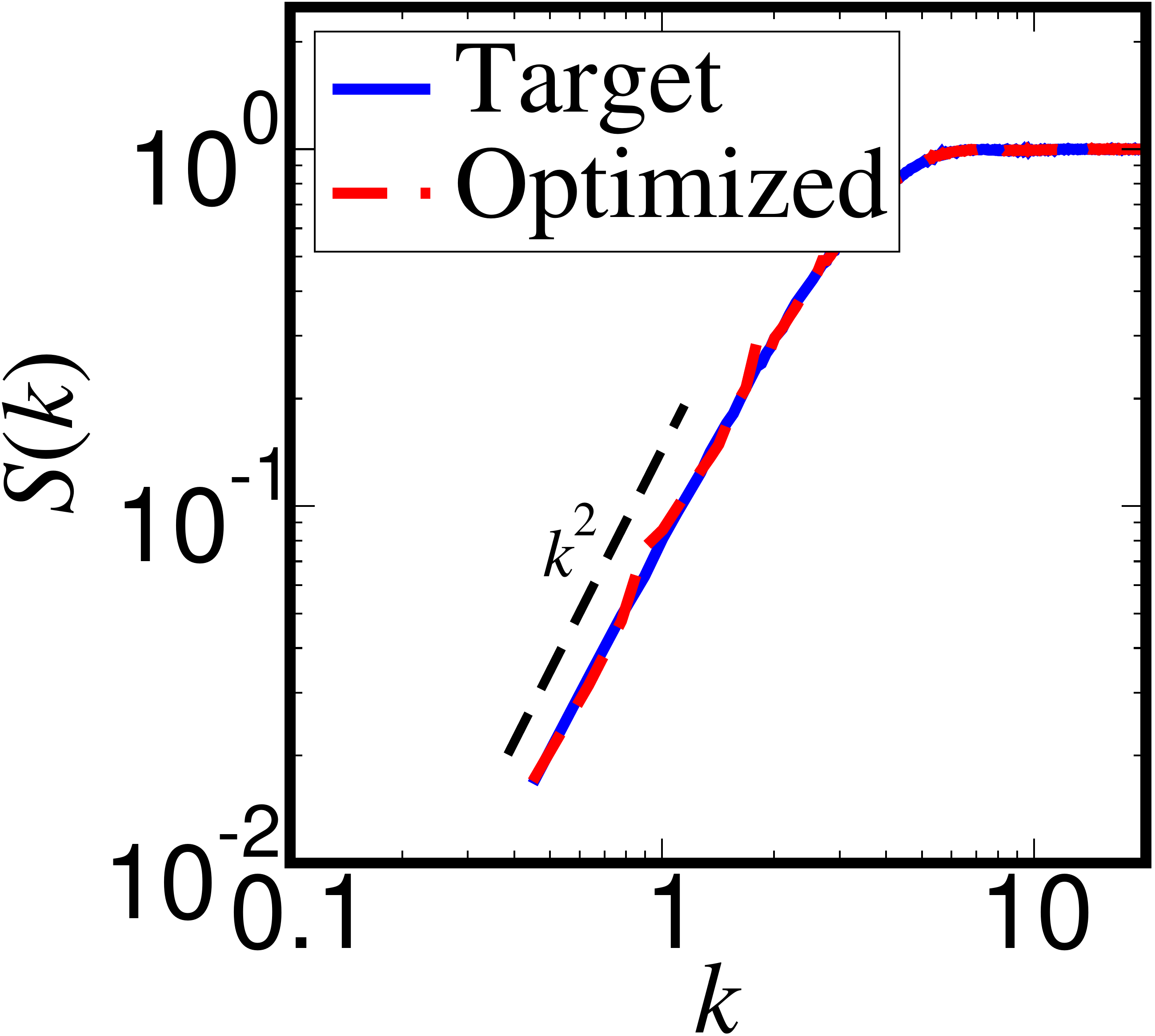}}
  \caption{(a) A portion of a representative 3D 2,744-particle configuration of a cloaked URL system. Only 512 particles are displayed. (b) A portion of a 3D configuration of a 2,744-particle system that is equilibrated under the optimized effective one- and two-body potential for the target 3D cloaked URL. Only 512 particles are displayed. (c) Optimized pair potential minus its long-ranged repulsive part $0.940/r$. (d) Targeted and optimized pair correlation functions with $N=9261$. Here we find that the $L_2$ function is $D_{g_2}=4.8\times 10^{-4}$. (e) Targeted and optimized structure factors with $N=9261$. Here we find the $L_2$ function is $D_{g_2}=7.5\times 10^{-4}$. The $L_2$-norm error is $\mathcal{E}=0.035$. 
  (f) Log-log plot of the targeted and optimized structure factors, showing their $k^2$ scaling behavior at small $k$. Subfigures (a)--(e) are reproduced from the ones first presented in Ref. \citenum{To22}.}
  \label{3DURL_basic}
\end{figure}

Figure \ref{3DURL_basic} shows the configurations, effective pair potential and pair statistics of the target and optimized systems for the 3D URL model, which have been previously determined and presented in Ref. \citenum{To22}.
Figures \ref{3DURL_basic}(a) and \ref{3DURL_basic}(b) that the target and the optimized systems configurations are visually indistinguishable. 
Figure \ref{3DURL_basic}(c) shows the short-ranged part of the effective potential, i.e. $v(r;\mathbf{a})-0.940/r$ against $r$. 
Figure \ref{3DURL_basic}(d)--(f) show $g_2(r)$ and $S(k)$ for the target and optimized systems with $N=9261$.
The pair statistics of the target cloaked URL is in excellent agreement with those of the optimized equilibrium system in both direct and Fourier space. 
The $L_2$ functions are $D_{g_2}=4.8\times 10^{-4}$ and  $D_{S}=7.5\times 10^{-4}$ and the $L_2$-norm error is $\mathcal{E}=0.035$, all of which are remarkably small.
For all models, the effective potentials accurately yield the desired target pair statistics for system sizes much larger than $N=500$ used in the optimization procedure. 
Thus, the nonequilibrium-equilibrium correspondences found in this study are robust to further increases in the system size.

\subsection{Higher-order statistics}
\label{sec:res_higher}
Since the targeted and optimized systems have essentially the same pair statistics, we expect that due to the aforementioned structural degeneracy, the dynamics leading to the nonequilibrium states is reflected in the differences in their higher-order statistics compared to the corresponding equilibrium states.
Here, we present the higher-order statistics of the nonequilibrium-equilibrium pairs, including $G_V(r)$ for all models and $g_3$ for the perfect glass and the critical absorbing state at specific small triangles. Note that $g_3$ and $g_4$ for the URL have been discussed in Ref. \citenum{Kl20}.

\begin{figure}[htp]
  \centering
  \subfloat[]{\includegraphics[width=26mm]{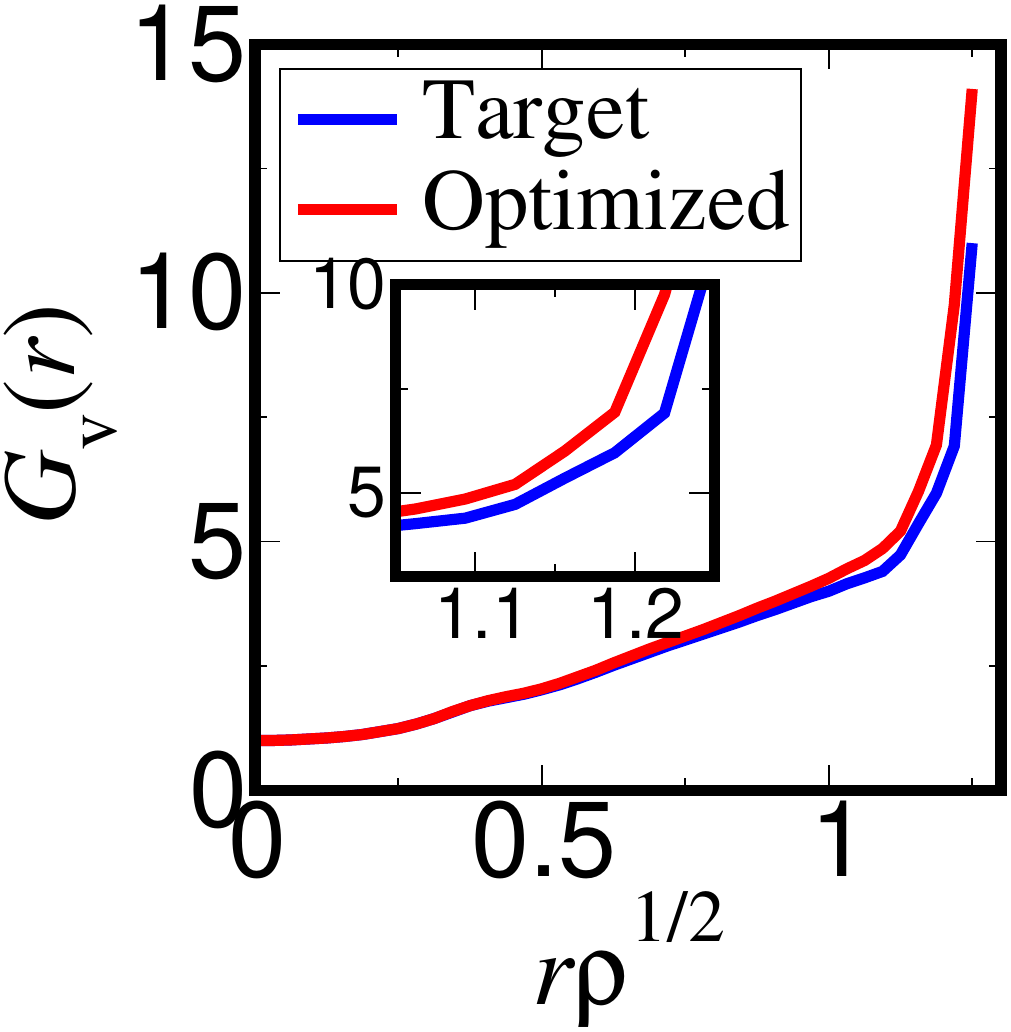}}
  \hspace{1mm}
  \subfloat[]{\includegraphics[width=26mm]{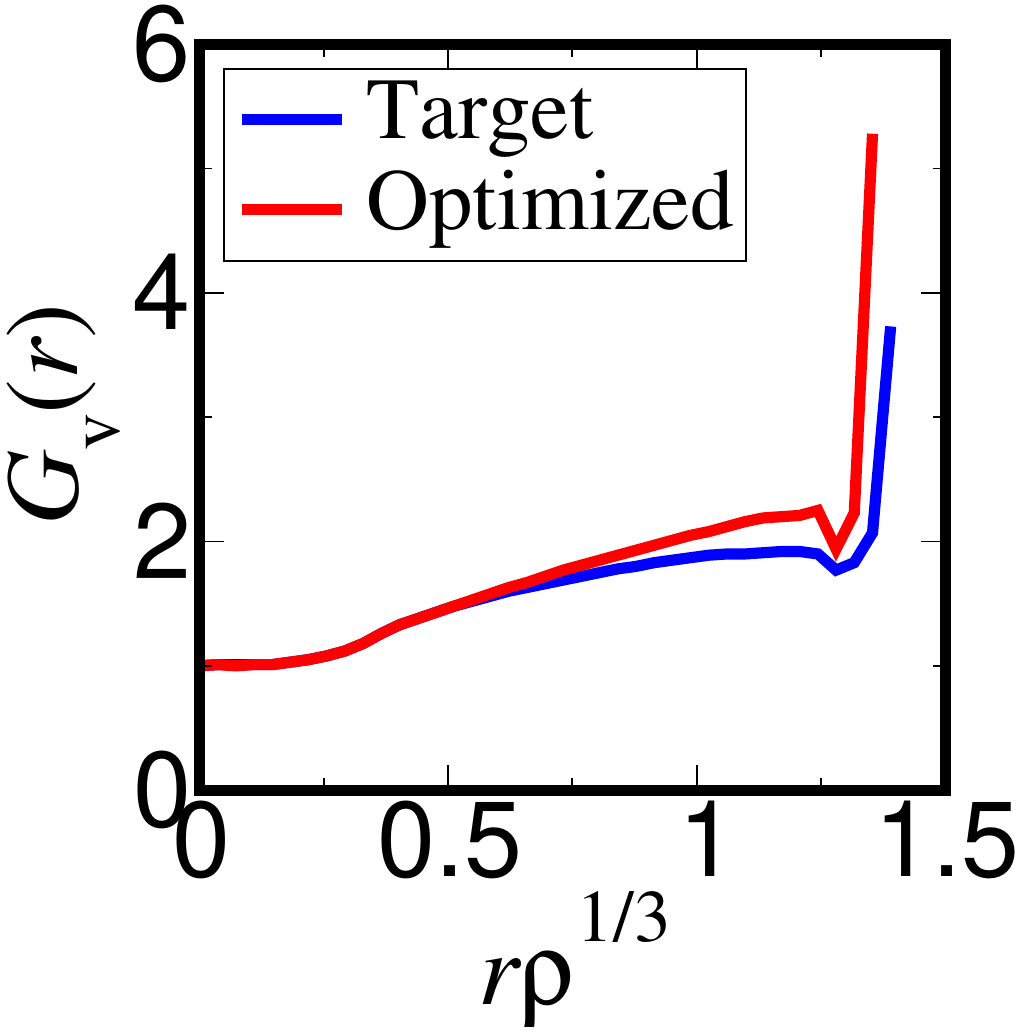}}
  \hspace{1mm}
  \subfloat[]{\includegraphics[width=26mm]{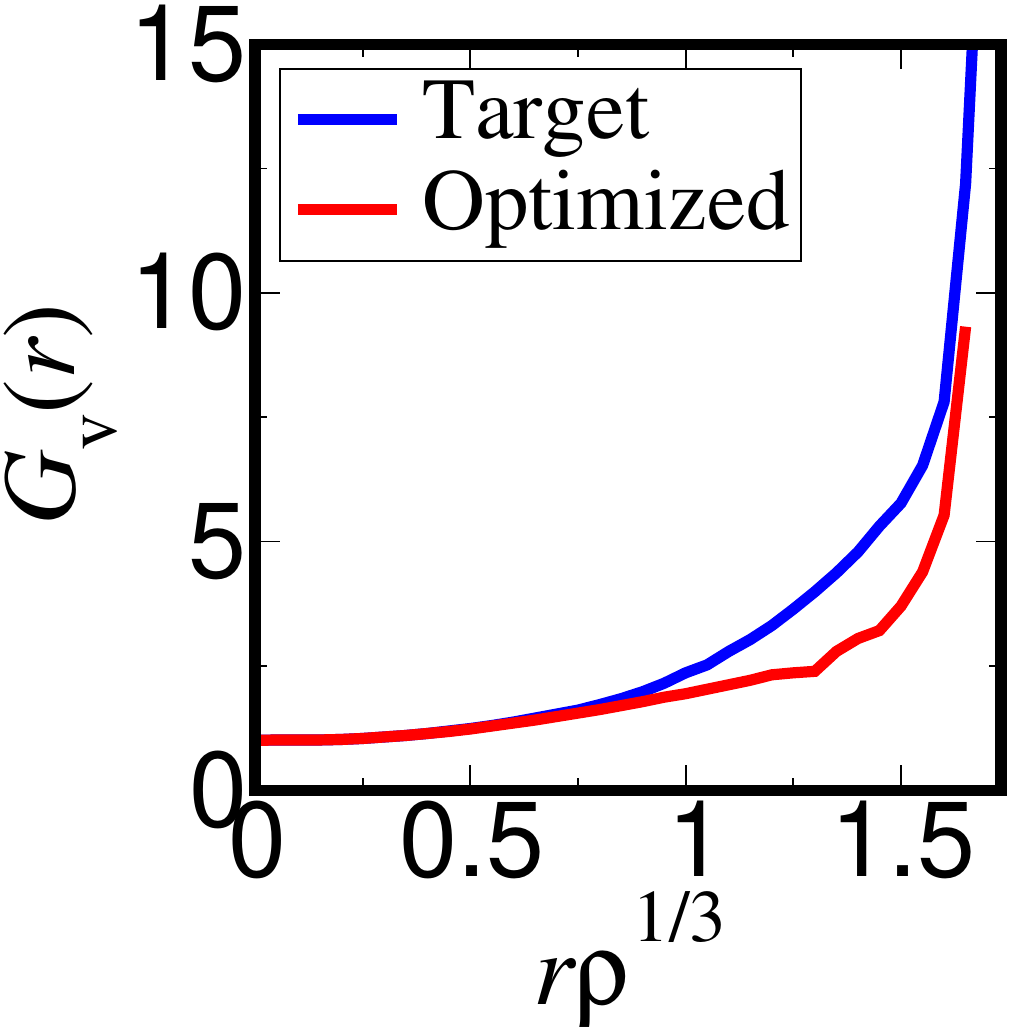}}
  \caption{Conditional ``hole'' probability density functions for three target and optimized systems. Note the different scales in the axes. (a) 2D perfect glass. The inset shows a magnified portion of the plots. (b) 3D critical absorbing-state model. (c) 3D cloaked URL.}
  \label{fig:gv}
\end{figure}

%\footnotetext[11]{We used $nN=2.4\times 10^6$, where $n$ is the number of configurations.}
In Appendix A, we describe a new and precise numerical algorithm to compute the hole probabilities for the nonequilibrium cloaked URLs in any space dimension $d$, which exploits the independence of particles in the URLs to achieve much higher accuracy than the standard method. \cite{Kl20}
Figure \ref{fig:gv}(a) and (b) show $G_V(r)$ of the nonequilibrium-equilibrium pairs for the 2D perfect glass, the 3D critical absorbing state and the 3D cloaked URL, respectively. For all three models, while the target and optimized $G_V(r)$ agree closely in the range where $r$ is below half of the characteristic length scale, they become increasingly different at larger $r$, revealing the distinctly different higher-order statistics between the nonequilibrium and equilibrium systems. Importantly, both target and optimized $G_V(r)$ for all systems studied exhibit steep rises at sufficiently large $r$, indicating that they possess the bounded-hole property. \cite{Zh17a,Gh18}

The forms of the effective potentials for the perfect glass and critical-absorbing state allow us to conclude that these models have the bounded-hole property, since the potentials contain hard cores (an effective hard core in the perfect-glass case), small but positive repulsions at intermediate $r$, and long-ranged repulsions at large $r$ that ensure hyperuniformity. Thus, particles of the resulting packings are well-separated from one another at all length scales. 
It is noteworthy that the bounded hole property applies for perfect glasses with any $\alpha>0$, because of the nature of the long-ranged two-, three- and four-body interactions.\cite{Zh16a} 
Indeed, we have also computed the hole probabilities for a different perfect glass with $\alpha=2$ and confirmed its bounded-hole property.

Using the technique described in Appendix A, we accurately determine that $E_V(r)$ for the target URL in the limit $r\rightarrow r_c^-$ exhibits a form first derived in Ref. \citenum{Mi20}, namely, $E_V(r)\sim (r_c-r)^\gamma$, where $\gamma=32$. 
Thus, Eq. (\ref{ev_gv}) gives
\begin{equation}
    G_V(r)\sim \frac{\gamma}{\rho s_1(r)(r_c-r)} \quad r\rightarrow r_c^-
\end{equation}
i.e., $G_V(r)$ has a pole of order one, as derived in Ref. \citenum{Mi20}. Compared to the SC lattice, for which $\gamma=d=3$, \cite{Mi20} the very large $\gamma$ value for the URL indicates that holes close to the critical-hole size are rarer in disordered systems, as noted in Ref. \citenum{Mi20}. Also note that unlike the other two models, $G_V(r)$ for the target URL is higher than the optimized equilibrium $G_V(r)$, suggesting that the target URL is locally more homogeneous than its equilibrium counterpart due to the underlying lattice. 

%The radial distance of the steep rise in $G_V(r)$ at $r\rho^{1/3}=1.35$ coincides with the radial distance at which the ``bump'' vanishes. Note that $G_V(r)$ for the target and optimized critical absorbing state possess local minima at $r\rho^{1/3}=1.3$, immediately before the sudden rise. These minima correspond to the length scale of void spaces that separate particle clusters. 

We propose the following scalar nonequilibrium index, $\Gamma$, that reflects differences in the higher-order correlation functions $g_3, g_4, \ldots$ for a nonequilibrium-equilibrium pair with matching pair statistics:
\begin{equation}
    \Gamma=\rho\int_{|\mathbf{r}|<r_{\text{max}}} \left[G_V(|\mathbf{r}|)-G_{V,T}(|\mathbf{r}|)\right]^2 d\mathbf{r},
    \label{Gamma}
\end{equation}
where $G_{V}$ and $G_{V,T}$ are conditional ``hole'' probability density distributions for the optimized and target systems, respectively, and $r_{\text{max}}$ is a cutoff radius corresponding to the largest hole radius detected in an ensemble of finite-size configurations.\footnote{If the target and optimized systems have different largest detected hole radii, $r_{\text{max}}$ is chosen to be the smaller of the two. 
The aforementioned difference is within $0.1\rho^{1/d}$ for all three models.} By definition, $G_V(r)$ in (\ref{Gamma}) must be computed at $kT/\varepsilon=1$, because it is only at this dimensionless temperature that the nonequilibrium and equilibrium pair statistics match. Note that $\Gamma$ is a purely \textit{static} nonequilibrium index, i.e., its computation does not require dynamic information. Importantly, for a statistically homogeneous \textit{equilibrium} target system under up to two-body interactions, one has $\Gamma=0$, because the optimized pair potential obtained via the inverse methodology (Sec. \ref{sec:meth}) must agree with the unique target-generating potential, yielding $G_V(r)=G_{V,T}(r)$ for all $r$.

Except for the target URL, whose $G_V(r)$ can be evaluated up to arbitrary precision, we determine $r_\text{max}$ and compute $G_V(r)$ using $500$ configurations with $N=2500$ for 2D systems and $N=9261$ for 3D systems.
Table \ref{tab:gamma} shows the value of $\Gamma$ for the three models shown in Fig. \ref{fig:gv}. The URL has a much larger $\Gamma$ value compared to those of the the other two systems. This high degree of ``nonequilibriumness'' is expected for the cloaked URL, since $g_4$ for the target system exhibits long-range order characteristic of the underlying lattice. \cite{Kl20} An equilibrium fluid cannot possess any long-range order.

\begin{table}[htp]
    \centering
    \caption{\ Values of the metric $\Gamma$ [Eq. (\ref{Gamma})] for the systems shown in Fig. \ref{fig:gv}}
    \begin{tabular}{l l}
    \hline
    System & $\Gamma$  \\
    \hline 
    2D perfect glass &  4.9 \\
    3D critical-absorbing state & 9.2 \\
    3D cloaked URL & 54 \\
    \hline
    \end{tabular}
    \label{tab:gamma}
\end{table}

\begin{figure}[htp]
  \centering
  \subfloat[]{\includegraphics[width=40mm]{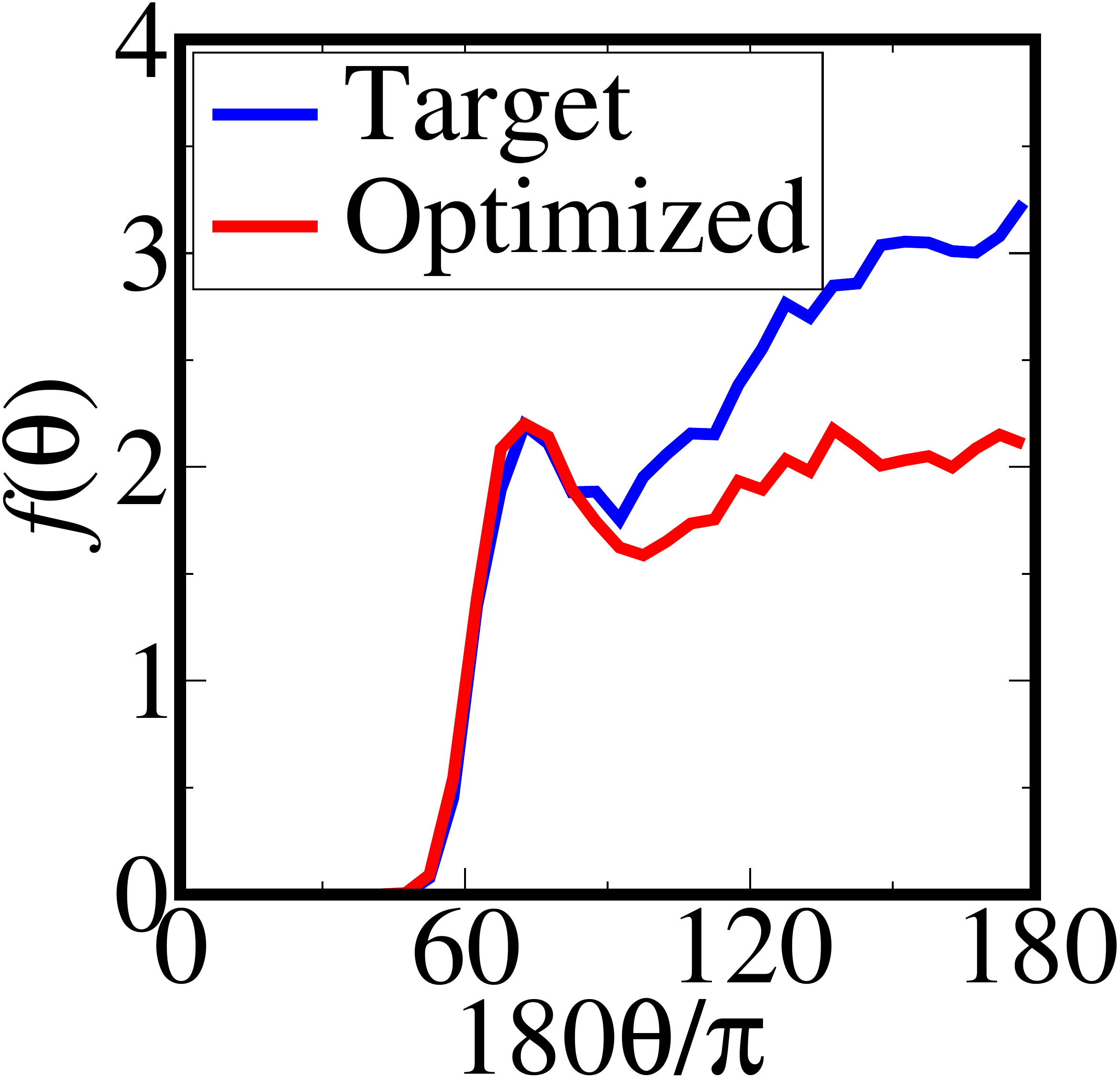}}
  \hspace{1mm}
  \subfloat[]{\includegraphics[width=40mm]{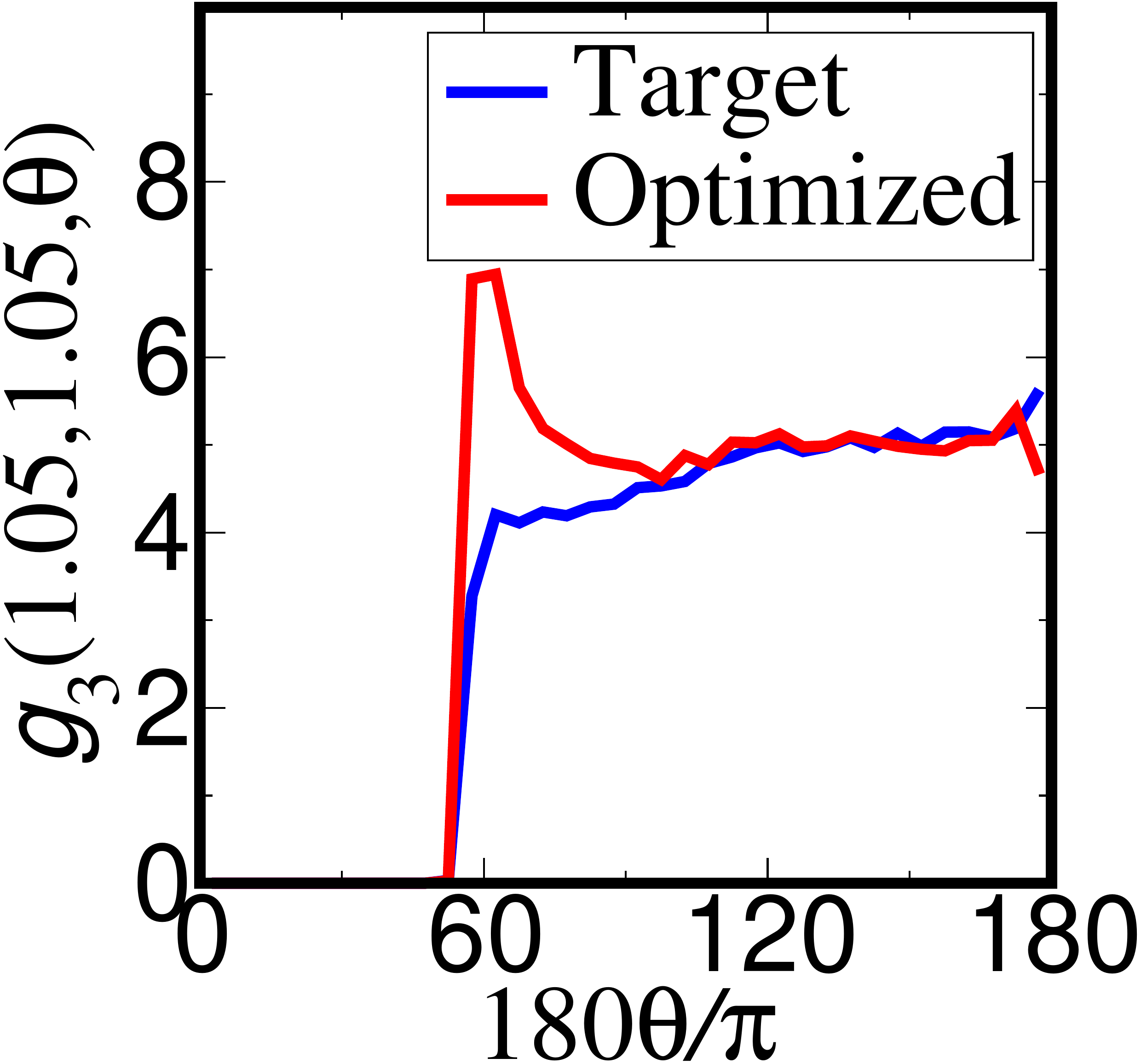}}
  \caption{(a) Plot of $f(\theta)=\int_{|\mathbf{r}_1|=0}^{1.15}\int_{|\mathbf{r}|_2=0}^{1.15} g_3(\mathbf{r}_1,\mathbf{r}_2,\theta)d\mathbf{r}_1d\mathbf{r}_2$ against $\theta$ for the target and optimized perfect glass systems. (b) Plot of $g_3(1.05,1.05,\theta)$ against $\theta$ for the target and optimized critical absorbing state systems.}
  \label{fig:g3}
\end{figure}

We also study the three-body statistics for the  perfect glass and the critical absorbing state at small triangles. In the case of the perfect glass, we compute the integral
\begin{equation}
    f(\theta)=\int_{|\mathbf{r}_1|=0}^{1.15}\int_{|\mathbf{r}_2|=0}^{1.15} g_3(r_1,r_2,\theta)d\mathbf{r}_1 d\mathbf{r}_2,
    \label{f}
\end{equation}
where $g_3(r_1,r_2,\theta)$ is given by (\ref{g3}). 
We evaluate $f(\theta)$ instead of $g_3$ at individual triangles, because much higher accuracy can be achieved for the former. 
Figure \ref{fig:g3}(a) depicts $f(\theta)$ for the target and optimized perfect glass systems. We find that compared to the equilibrium system, the target perfect glass contains 50\% more nearly linear small triangles with $\theta>170^\circ$. This difference is likely due to the actual three and four-body potential in the target perfect-glass system. \cite{Zh16a}

In the case of the critical-absorbing-state model, we compute $g_3$ at individual triangles, as it can be evaluated with high accuracy. Figure \ref{fig:g3}(b) shows $g_3(r_1,r_2,\theta)$ for small isosceles triangles with $r_1=r_2=1.05$ for the target and optimized critical absorbing state, from which it is clear that the target system significantly suppresses the formation of small triangles that are nearly equilateral. Indeed, their difference of $g_3$ for small equilateral triangles is 90\%. This is due to the specific dynamics during the random organization process, where particles are displaced to avoid collisions. \cite{Co08}  Dense clusters that contain many small equilateral triangles are less likely to be found in the nonequilibrium state, since collisions are more frequent within such clusters. On the other hand, the equilibrium state forms more dense clusters due to the sharp well of $v(r)$ at the hard-sphere diameter.

\subsection{Inherent structures}
\label{sec:res_is}
To study the effect of quenching on the hyperuniform equilibrium systems, we ascertain the inherent structures of the perfect-glass and critical-absorbing-state potentials by finding deep local energy minima via the low-storage BFGS algorithm, \cite{Liu89} starting from equilibrium initial configurations with $N=1000$ for $d=2$ and $N=1728$ for $d=3$, and averaging over 200 configurations. We find that the energy per particle $\overline{E}_{\text{IS}}$ of these structures are narrowly distributed above the ground-state energies; see Appendix C for details.
We compute the inherent-structure pair statistics by averaging over configurations whose energies are in within two standard deviations around the mean value of $\overline{E}_{\text{IS}}$.

\begin{figure}[htp]
  \centering
  \subfloat[]{\label{pgAlpha1IS_snap}\includegraphics[width=38mm]{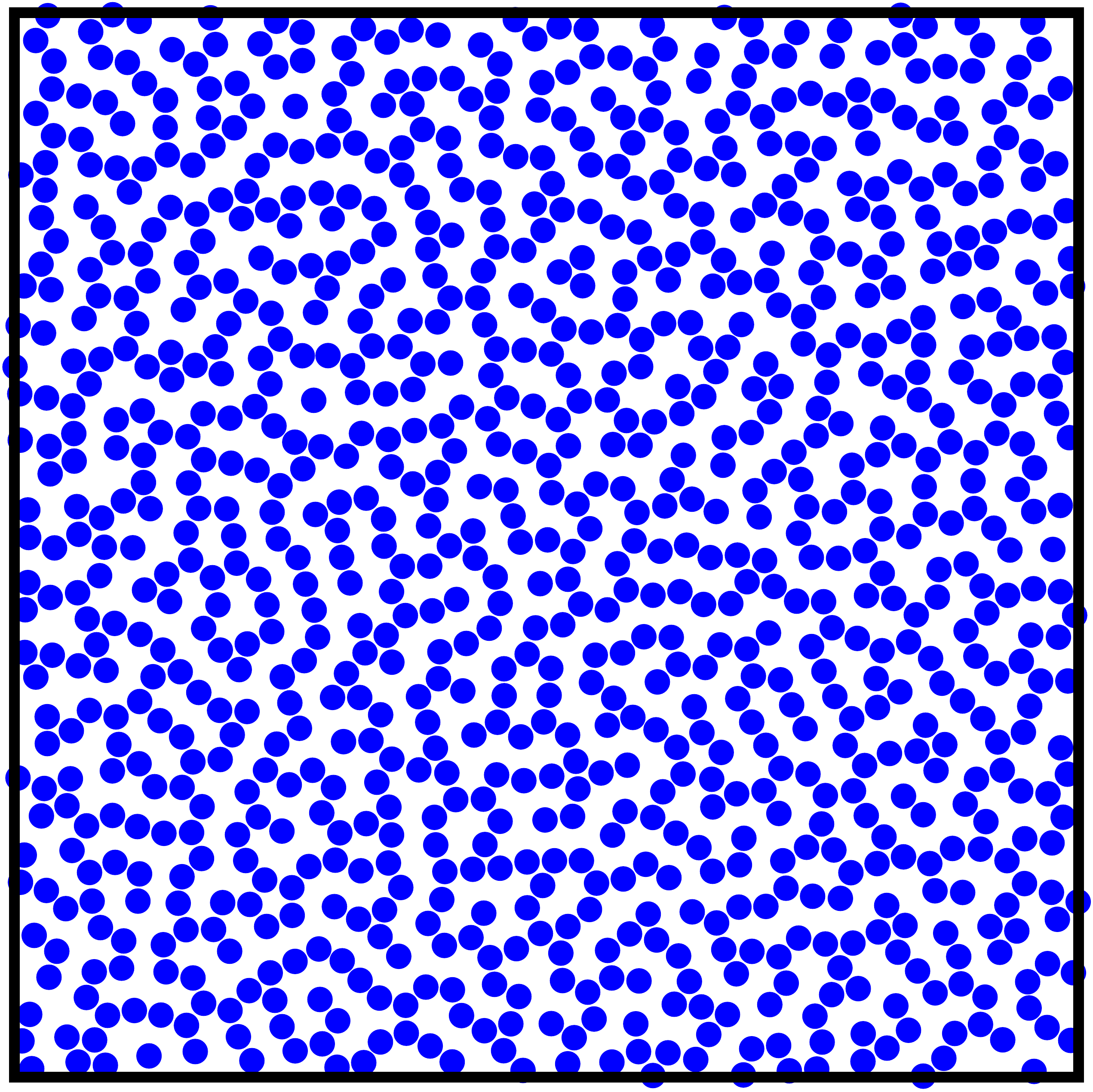}}
  \hspace{1mm}
  \subfloat[]{\label{pgAlpha1IS_g2}\includegraphics[width=40mm]{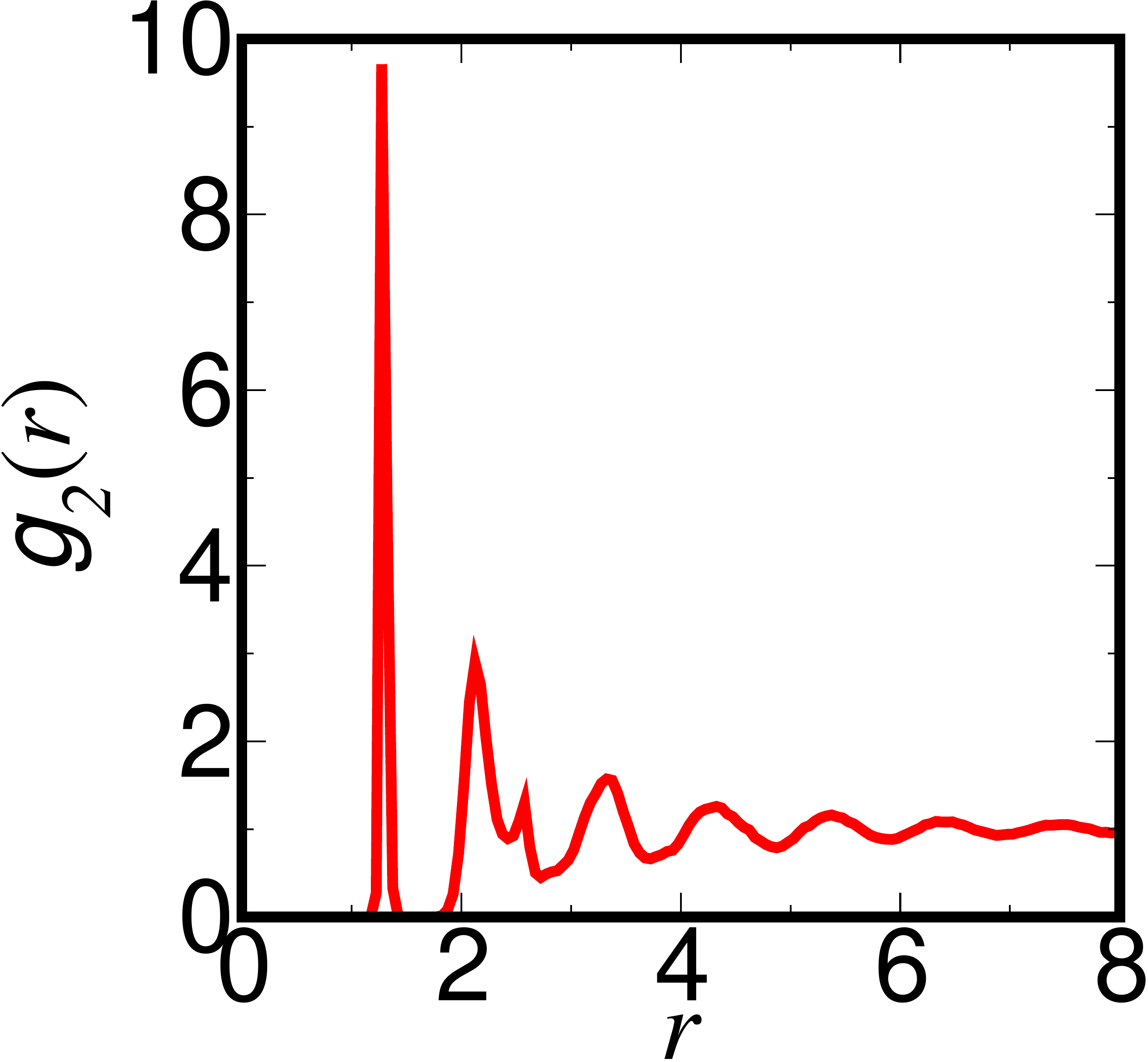}}
  
  \subfloat[]{\label{pgAlpha1IS_S}\includegraphics[width=38mm]{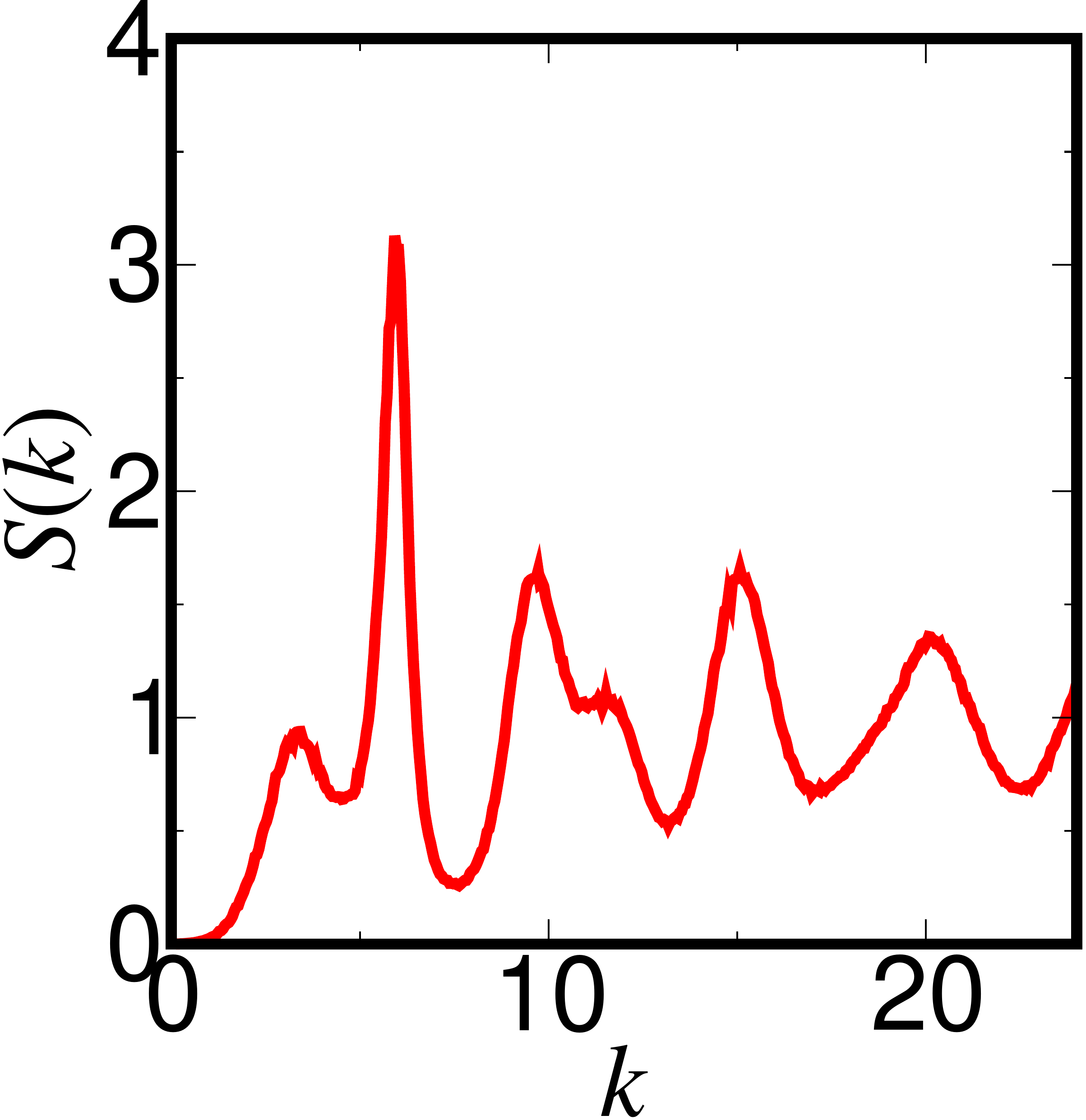}}
  \hspace{1mm}
  \subfloat[]{\label{pgAlpha1IS_SLog}\includegraphics[width=40mm]{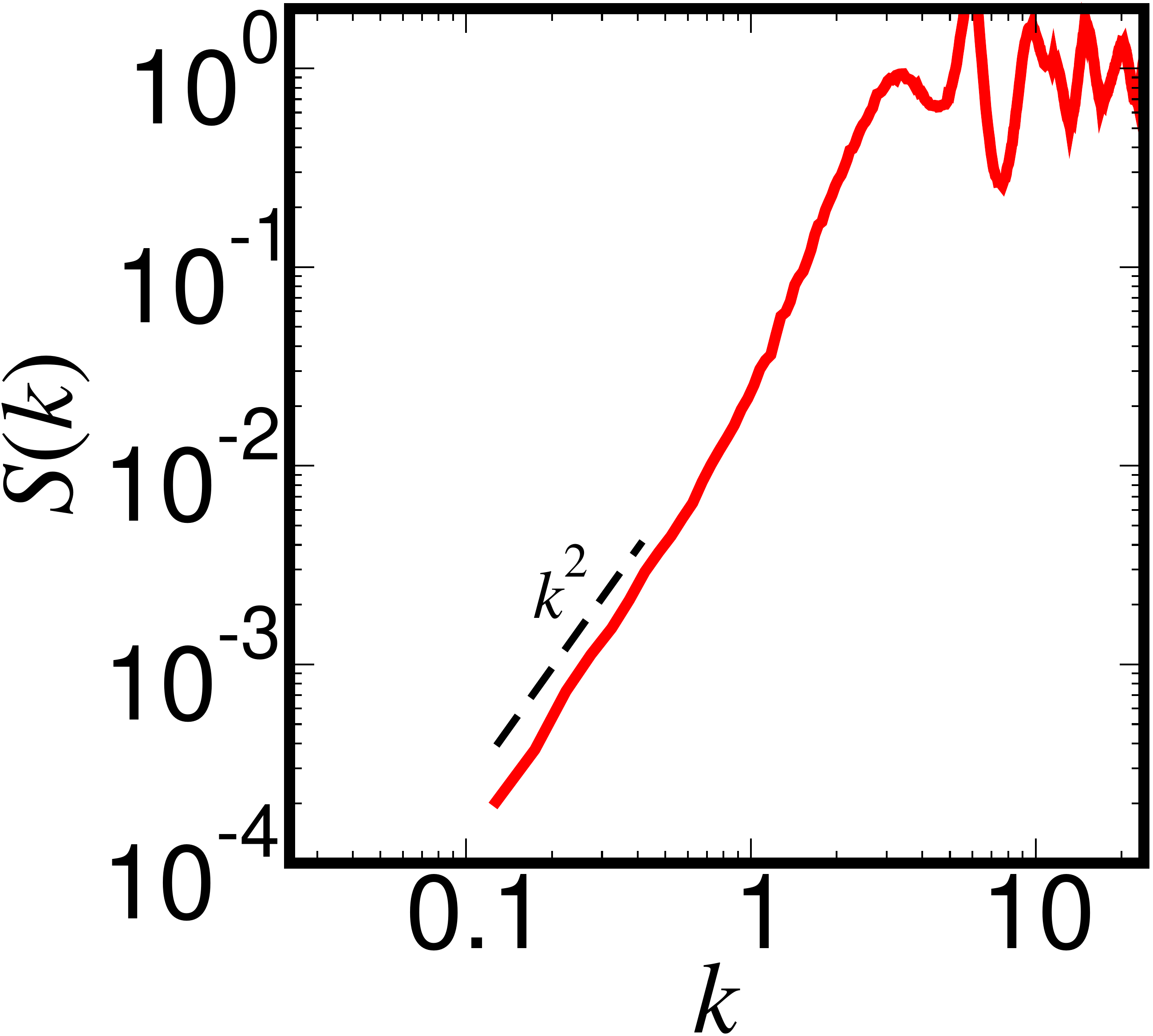}}
  \caption{(a) A 1,000-particle configuration of an inherent structure for the effective potential for the perfect glass. (b) Pair correlation function of the inherent structure. (c) Structure factor of the inherent structure. 
  (d) Log-log plot of the structure factor for the inherent structure, showing the $k^2$ scaling at small $k$.}
  \label{fig:pgAlpha1IS}
\end{figure}

Figure \ref{fig:pgAlpha1IS}(a) shows a disordered inherent structure for the perfect-glass effective potential, which contains chains of particles separated by ``channels'' with a well-defined width. The first three peaks of the associated $g_2(r)$ [Fig. \ref{fig:pgAlpha1IS}(b)] correspond to the nearest-neighbor distance, the channel width, and the second-nearest neighbor distance along the chains, respectively. The first and second peaks occur at $r=1.27$ and $r=2.20$, respectively, which are close to the locations of the first and second minima of the short-ranged part of $v(r)$.
Figure \ref{fig:pgAlpha1IS}(c)  depicts $S(k)$ for the inherent structures. 

\begin{figure}[htp]
  \centering
  \subfloat[]{\label{randOrg3DIS_snap}\includegraphics[width=40mm]{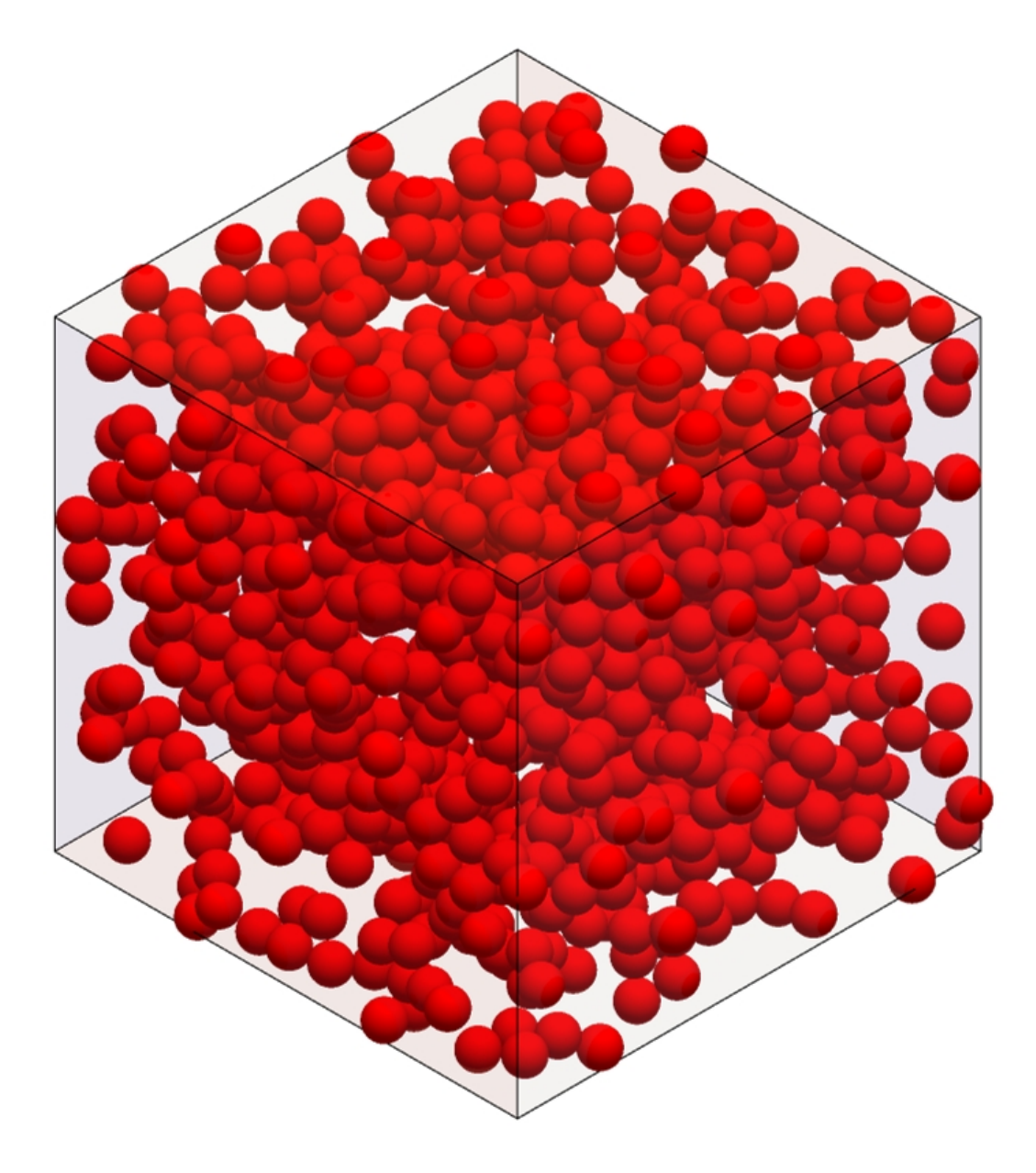}}
  \hspace{1mm}
  \subfloat[]{\label{randOrg3DIS_g2}\includegraphics[width=38mm]{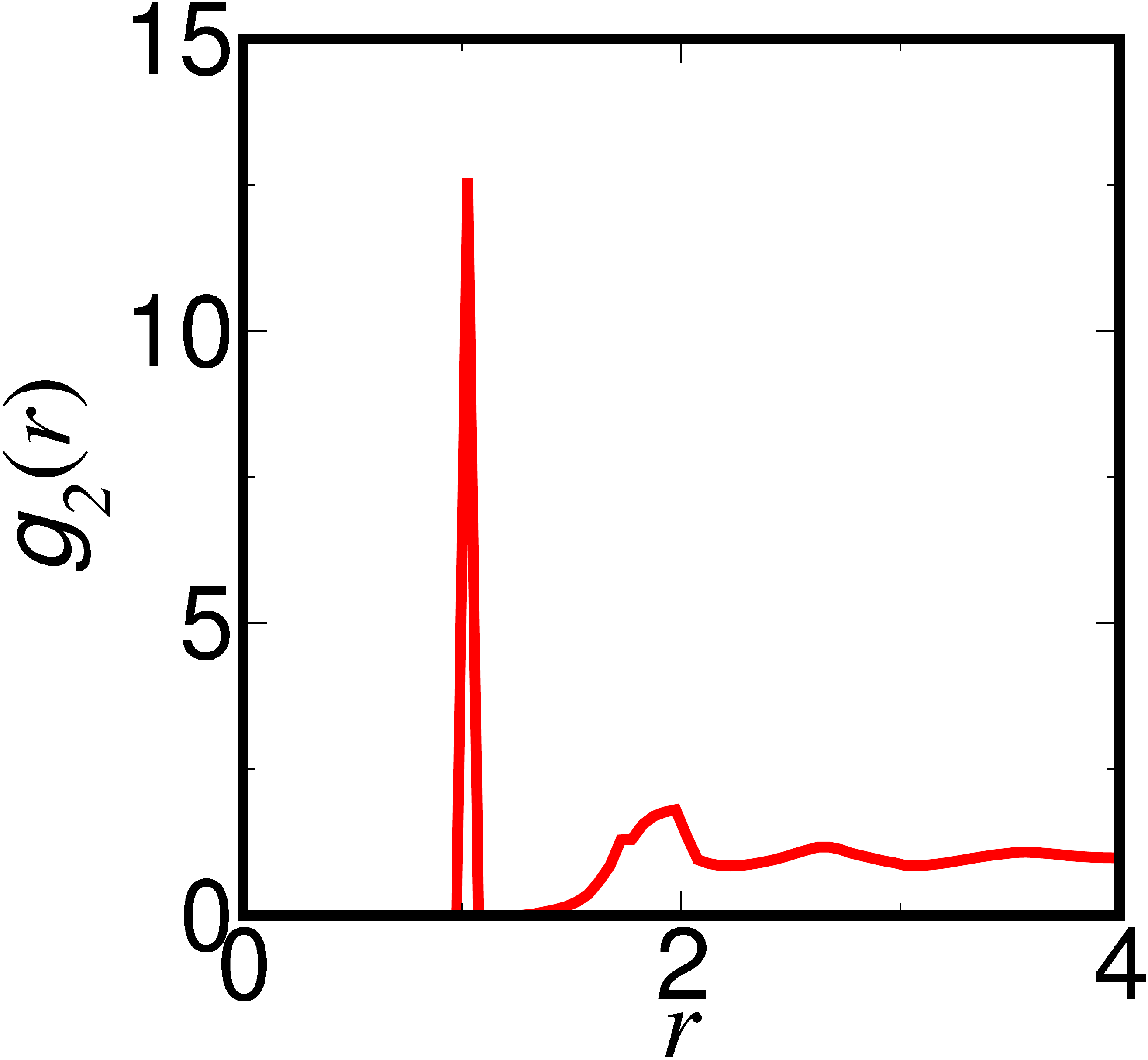}}

  \subfloat[]{\label{randOrg3DIS_S}\includegraphics[width=38mm]{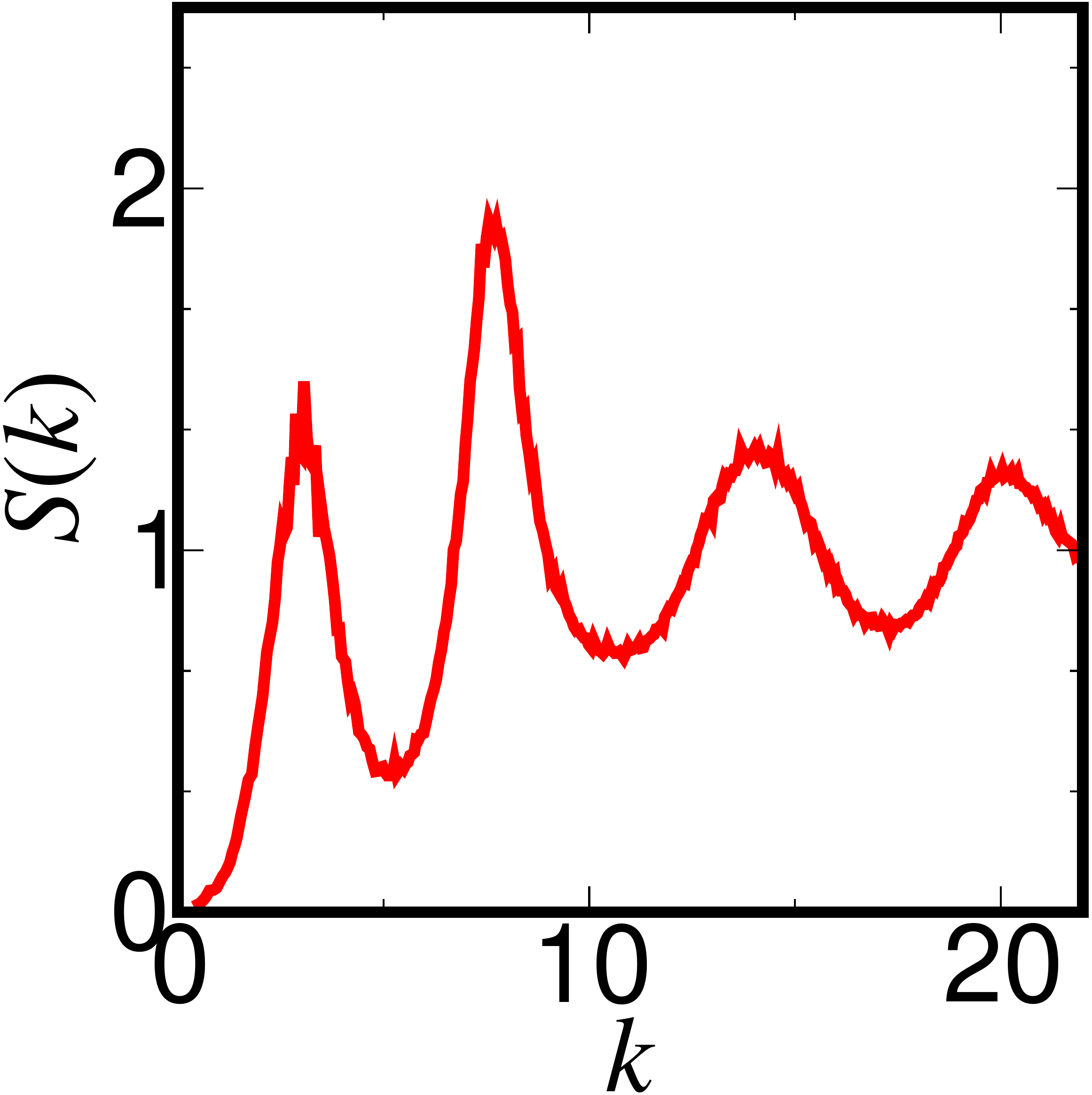}}
  \hspace{1mm}
  \subfloat[]{\label{randOrg3DIS_SLog}\includegraphics[width=40mm]{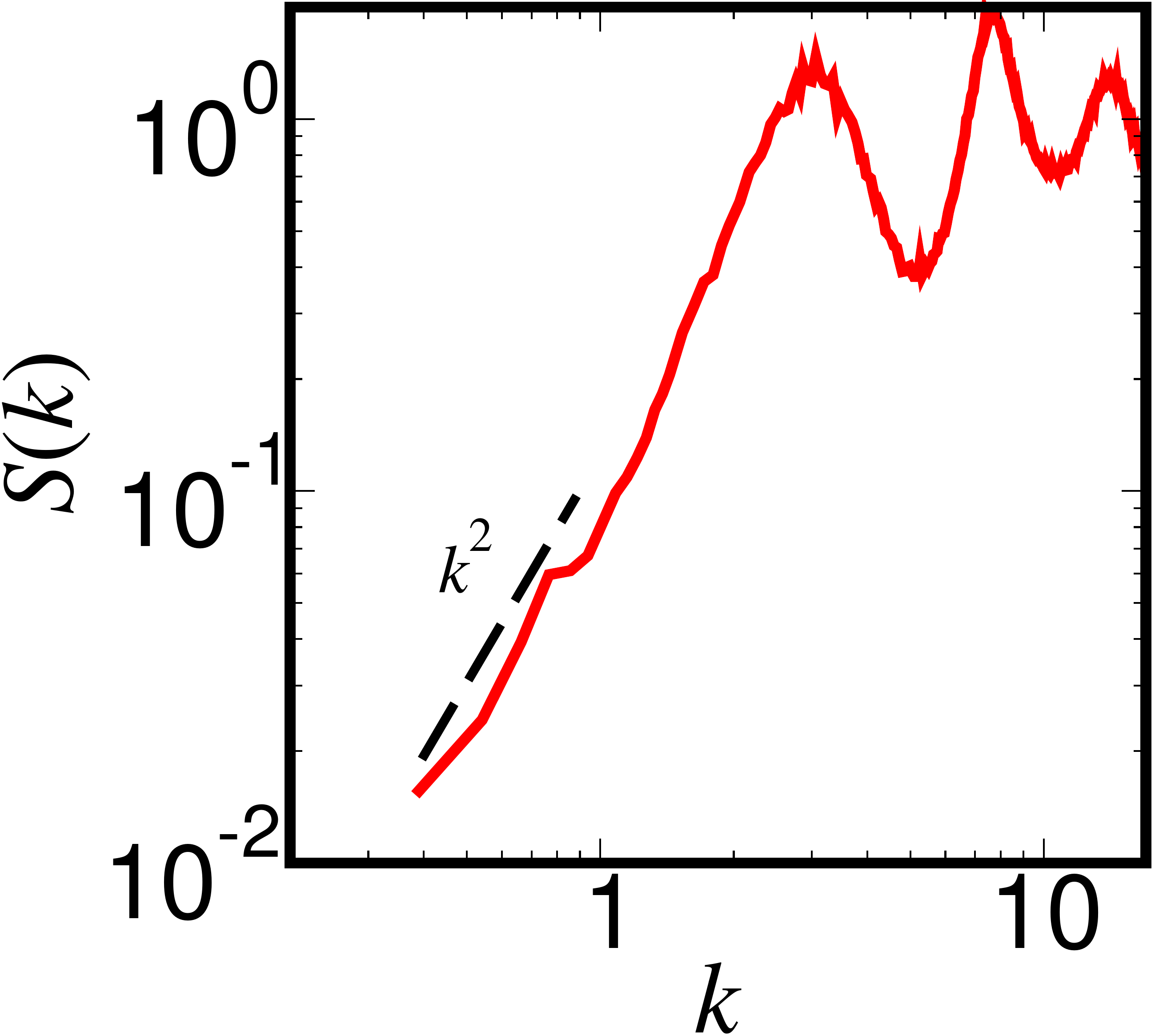}}
  \caption{(a) A 1,000-particle configuration of an inherent structure for the effective potential for the 3D critical-absorbing state. (b) Pair correlation function of the inherent structure with $N=1728$. (c) Structure factor of the inherent structure with $N=1728$. (d) Log-log plot of the structure factor for the inherent structure, showing the $k^2$ scaling at small $k$.}
  \label{fig:randOrgIS}
\end{figure}

To extract the small-$k$ behaviors of the structure factors, we use the concept of diffusion spreadability introduced in Ref. \citenum{To18d}, defined as follows. 
Consider the time-dependent problem of mass transfer of a solute between two phases and assume that the solute is initially distributed in one phase (phase 2) and absent from the other (phase 1). 
The spreadability $\mathcal{S}(t)$ is the fraction of total solute present in phase 1 as a function of time.
For sphere packings, which include the perfect glass and the critical absorbing state, $\mathcal{S}(t)$ can be computed from simulated $S(k)$ via the Fourier-space expression derived in Ref. \citenum{To18d}. 
Recently, Wang and Torquato \cite{Wa22a} introduced an algorithm that efficiently and accurately extracts the exponent $\alpha$ from numerical data of $\mathcal{S}(t)$. 
The algorithm extracts $\alpha$ as well as a ``set-in'' time $t_S$ of the asymptotic behavior of $\mathcal{S}(t)$ by solving the following relations via a predictor-corrector procedure
\begin{equation}
\begin{split}
    |\mathcal{S}(t)-\mathcal{S}_l(t;\alpha)|=\epsilon, \quad t=t_S,\\
     |\mathcal{S}(t)-\mathcal{S}_l(t;\alpha)|<\epsilon, \quad t>t_S
\end{split}
\end{equation}
where $\epsilon>0$ is a convergence criterion and $\mathcal{S}_l(t;\alpha)$ is the large-$t$ spreadability approximant. 
For nonstealthy media, one has $t^{-(d+\alpha)/2}$. \cite{To18d}
It has been shown that the aforementioned algorithm is more robust to simulation noise than a direct numerical fit of $S(k)$ at small $k$, and accurately extracts $\alpha$ with errors less than 1\% for a wide range of models. \cite{Wa22a} 
We find that the inherent structures are hyperuniform with $\alpha=2.0$.
Figure \ref{fig:pgAlpha1IS}(d) depicts $S(k)$ for the inherent structures on a log-log scale, showing clearly the $k^2$ behavior at small $k$.
Thus, the inherent structures of a stronger form of hyperuniformity (class I) than that of the target structure. 
Figure \ref{fig:randOrgIS} shows a disordered inherent structure for the critical-absorbing-state potential, as well as the inherent-structure pair statistics, indicating that it is again hyperuniform with $\alpha=2$. By contrast, $\alpha$ for the URL potential \cite{To22} dramatically increases from 2 at unit temperature to infinity upon quenching, as the inherent structure is the SC lattice. Thus, in all three cases, the inherent structures are of higher forms of hyperuniformity than the target models. 

To verify that the increase of $\alpha$ upon quenching is robust to the system size, we also computed the inherent structures for the critical-absorbing state with $N=512, 729$ and $1000$, in addition to $N=1728$. 
We observed that while the structure factors for these system sizes are slightly different due to finite-size effect, they all exhibit $k^2$ scaling behaviors at small $k$, indicating that the inherent structures possess a large hyperuniformity scaling regime \cite{To21c} that grows with $N$. 

\section{Conclusions and discussion}
\label{sec:conc}
We have determined equilibrium systems with effective one- and two-body interactions that realize the pair statistics of three hyperuniform nonequilibrium models of recent interest: a 2D perfect glass, a 3D critical-absorbing state and a 3D cloaked URL, which lends further support to the Zhang-Torquato conjecture. 
In the case of the perfect glass, it is noteworthy that we are able to mimic systems with up to four-body interactions through effective one-body and isotropic two-body interactions. 
This finding offers promise to determining effective one- and two-body interactions that mimic systems with directional interactions, such as amorphous graphene \cite{Ch21b} and amorphous silicon \cite{He13}, governed by up to three- and higher-body interactions. \cite{Li10, St85b} We have shown that all three models considered in this study have the bounded-hole property.

The volume integral of the squared differences of $G_V(r)$ between such nonequilibrium and equilibrium systems with identical pair statistics enable us to define a nonequilibrium index $\Gamma$ [Eq. (\ref{Gamma})], which can be applied to measure the ``nonequilibriumness'' in many systems of practical interest, including supercooled liquids, \cite{Ma13a} defects, \cite{Ka01b} self-propelling particles \cite{Hu21, Zh22} and biological tissues. \cite{Bi16} 
We remark that a different static nonequilibrium index $X$ has been proposed previously, which is based on the deviation of $S(0)$ from $\rho \kappa_T k_B T$, where $\kappa_T$ is the isothermal compressibility. \cite{Ma13a} 
In comparison, $\Gamma$ extracts the nonequilibriumness from structural information alone, and can be applied in situations where $\kappa_T$ (or the pressure) is not readily available. 
We also introduced a precise numerical algorithm to compute $E_V(r)$ for the cloaked URL, and applied it to show that holes near the maximum hole size in the URL are much rarer than those in the underlying SC lattice.

Remarkably, our effective potentials yield hyperuniform deep local energy minima that are of stronger forms of hyperuniformity (measured by the class or larger $\alpha$  exponents) relative to the equilibrium states at unit temperature. 
This behavior is expected for pair potentials characterized by steep short-range repulsions, since the local minima can be regarded to be states with effectively low temperatures relative to the initial higher-temperature fluid \cite{Zh21b} and hence the structure factor $S(k)$ of the quench at low wavenumbers reflects stronger hyperuniformity.
By contrast, the same reasoning leads to the extended proposition that a quench of an initially nonhyperuniform fluid state with such interactions to deep local energy minima have a $S(0)$ that drops but generally not to zero, i.e., it strictly remains nonhyperuniform, as shown for three well-known models in Appendix C.
%Since an equilibrium system at positive $T$ can be regarded as a thermally excited state of a corresponding inherent structure \cite{St82a,St82}, our observation that $\alpha$ increases upon quenching shows that thermal excitation of hyperuniform systems with long-ranged interactions could degrade, but not destroy, hyperuniformity \cite{Ki18a}.

A promising avenue for future research is the determination of thermodynamic and dynamic properties associated with the effective potentials, including ground states, entropies, free energies, phase diagrams, \cite{Mi04, Pa09} and their glass formations. For instance, it has been shown that perfect glasses at $T=0$ are remarkably zero-entropy states, \cite{Zh17b} in contradistinction to normal glasses that are metastable with respect to crystals, and to disordered stealthy ground states that possess large extensive entropies. \cite{To15} On the other hand, the equivalent equilibrium system for a perfect glass has positive entropy, and an effective pair potential [Fig. \ref{fig:pgAlpha1}(c)] yields a crystalline ground state, as shown in Appendix C. Furthermore, we note that the equivalent nonequilibrium and equilibrium systems possess the same two-body contribution to the excess entropy per particle due to their identical pair statistics. \cite{Ha86} 
However, this excess two-body entropy can be significantly different from the real entropy even for simple equilibrium liquids. \cite{Lo17, Zu20}
Thus, it is a fascinating problem to study their higher-order correlation contributions to the entropy, which is expected to reveal crucial dynamical information.

Finally, we stress that the effective potentials yielding hyperuniform states at positive $T$ could enable one to produce tunable hyperuniform materials. Due to the long-ranged nature of these potentials (\ref{v_long_hu}), one can use $\rho$ and $T$ as tuning parameters to generate equilibrium hyperuniform structures whose exponents $\alpha$ are dictated by (\ref{v_long_hu}), or to generate stronger hyperuniform forms via their inherent structures. 
While it is challenging to achieve long-ranged interactions in the laboratory, one could experimentally reproduce the effective potentials over some finite but large range of $r$ to fabricate \textit{effectively hyperuniform} states, i.e., states with very small but nonvanishing $S(0)$. Subsequently, the deviation of such systems from perfect hyperuniformity can be characterized via the various quantitative measures described in Ref. \citenum{To21c}.

\section*{Conflicts of interest}
There are no conflicts to declare.

\section*{Acknowledgements}
The authors gratefully acknowledge the support of the National Science Foundation under Award No. CBET-2133179. S.T. thanks the Institute for Advanced Study for their hospitality during his sabbatical leave there.

\section*{Appendix}
\section*{A Precise numerical algorithm to compute hole probabilities for URLs}
\label{sec:url}
Here, we describe the standard method of computing the void-exclusion probability or ``hole'' probability function $E_V(r)$ for a many-body system. 
For systems with the bounded hole property, sampling holes with radii close to $r_c$ is computationally challenging. 
To precisely determine the behaviors of $E_V(r)$ and $G_V(r)$ as $r\rightarrow r_c^-$, we introduce an improved algorithm to compute $E_V(r)$ for URLs in all dimensions with any perturbation strength $b$ that is much more accurate than the standard method. 
Although this improved method only applies to URLs, we will see in Sec. \ref{sec:res_higher} that the results for the hole probabilities of this model confirms the general trend that holes close to the critical-hole size are rarer in disordered systems than in ordered systems. \cite{Mi20}
We first describe the algorithm specific to the 3D cloaked URL (i.e., for $d=3$ and $b=1$), and then generalize it to other $d$ and $b$.

In the standard method, $N_t$ test particles are randomly placed in a $N$-particle configuration $\mathbf{r}^N=\{\mathbf{r}_1,\dots \mathbf{r}_N\}$ under periodic boundary conditions. For each test particle $\mathbf{x}$, one computes the distance $d_{\mathbf{x}}$ from $\mathbf{x}$ to its nearest real particle $\mathbf{r}_i$. The function $E_V(r)$ is estimated to be
\begin{equation}
E_V(r)=\frac{|\{\mathbf{x}:d_{\mathbf{x}}>r\}|}{N_t}.
\end{equation}
The result is then averaged over $n$ configurations. Using $nN=5\times 10^7$, $N_t/N=10^5$, this method yields $E_V(r)$ for the target URL in the range $r\rho^{1/3}<1.3$ with errors $<5\%$. However, computing $E_V(r)$ at larger $r$ requires sampling rare events with probabilities smaller than $10^{-15}$, which is computationally challenging. \cite{Mi20} Thus, we exploit the independence of the perturbed lattice points to devise a more accurate method to compute $E_V(r)$.

As described in Sec. \ref{sec:models_url}, the cloaked URL is obtained by perturbing each lattice point in the SC lattice $\mathbb{Z}^3$ by a random vector uniformly distributed on $[-1/2,1/2)^3$. For $i,j,k\in \mathbb{Z}$, let $C_{ijk}=[i-1/2,i+1/2)\times[j-1/2,j+1/2)\times[k-1/2,k+1/2)$, i.e., the cubic region accessible to the perturbed lattice point originally at $(i,j,k)$. Let $\mathbf{x}$ be the position of a test particle and $\mathcal{B}(\mathbf{x};r)$ be the spherical region of radius $r$ centered at $\mathbf{x}$. 

Since the lattice points are perturbed independently, the probability $p(\mathbf{x},r)$ that no perturbed lattice point is found in $\mathcal{B}(\mathbf{x};r)$ is given by the product of the probabilities that each perturbed point is not in $\mathcal{B}(\mathbf{x};r)$, i.e.,
\begin{equation}
    p(\mathbf{x},r)=\prod_{ijk}|C_{ijk}\setminus\mathcal{B}(\mathbf{x};r)|,
    \label{pqr}
\end{equation}
where $|C_{ijk}\setminus\mathcal{B}(\mathbf{x};r)|$ is the volume of the set difference between $C_{ijk}$ and $\mathcal{B}(\mathbf{x};r)$. Because the probability distribution of the position of the test particle is uniform, we have
\begin{equation}
    E_V(r)=\langle p(\mathbf{x},r)\rangle_{\mathbf{x}\in \mathbb{R}^3}=\int_{C_{000}} p(\mathbf{x},r) d\mathbf{x},
    \label{Ev_URL}
\end{equation}
where the second equality follows from the fact that all $C_{ijk}$ are equivalent in a URL. In what follows, we let $\mathbf{x}\in C_{000}$.

If $r\geq\sqrt{3}$, the region $\mathcal{B}(\mathbf{x};r)$ covers $C_{000}$, i.e., $|C_{000}\setminus\mathcal{B}(\mathbf{x};r)|=0$ for any $\mathbf{x}\in C_{000}$; It follows from Eqs. (\ref{pqr}) and (\ref{Ev_URL}) that $E_V(r)=0$ for $r\geq \sqrt{3}$. On the other hand, if $0\leq r<\sqrt{3}$, then $p(\mathbf{x},r)\neq 0$ for $\mathbf{x}=(-1/2,-1/2,-1/2)^T$, and thus $E_V(r)$ is nonzero. Therefore, the maximum hole radius in the 3D cloaked URL is given by $r_c=\sqrt{3}$. 

For $r<r_c$, Eq. (\ref{pqr}) reduces to 
\begin{equation}
    p(\mathbf{x},r)=\prod_{|i|\leq 2, |j|\leq 2, |k|\leq 2}|C_{ijk}\setminus\mathcal{B}(\mathbf{x};r)|.
    \label{pqr_reduced}
\end{equation}
The factors in Eq. (\ref{pqr}) that are not included Eq. (\ref{pqr_reduced}) are equal to unity because $\mathcal{B}(\mathbf{x};r)$ are disjoint from their corresponding $C_{ijk}$. The 125 factors in Eq. (\ref{pqr_reduced}) can be evaluated via Monte Carlo integration technique. \cite{Pr86} We randomly place $N_s=10^6$ sample points in each $C_{ijk}$ region and estimate $|C_{ijk}\setminus\mathcal{B}(\mathbf{x};r)|$ to be the fraction of the number of sample points not found in $\mathcal{B}(\mathbf{x};r)$. 
We compute $p(\mathbf{x},r)$ at all $\mathbf{x}$ vectors on a $M\times M \times M$ regular mesh on $C_{000}$, where $M$ is a positive integer. The mesh is chosen to be finer (i.e., $M$ larger) with increasing $r$ to ensure that there are at least 1,600 $\mathbf{x}$ vectors on the support of $p(\mathbf{x},r)$. 
Subsequently, we numerically evaluate $E_V(r)$ [Eq. (\ref{Ev_URL})] using a Gaussian quadrature of order 3. \cite{Pr86} This method yields errors of $E_V(r)$ on the order of 2\% for $1.6\leq r<r_c$, whereas the errors using the standard method is on the order of 100\% in this range. Note that the errors can be further reduced by increasing $N_s$.

To generalize the algorithm to other $d$ and $b$, one simply replaces $C_{ijk}$ by $C_{i_1, ..., i_d}=[i_1-b/2,i_1+b/2)\times\dots \times[i_d-b/2,i_d+b/2)$, which is the hypercubic region accessible to each perturbed lattice point originally at $(i_1, \dots, i_d)\in \mathbb{Z}^d$ in the $d$-dimensional hypercubic lattice. Equations (\ref{Ev_URL}) and (\ref{pqr}) will then be replaced by
    \begin{equation}
    E_V(r)=\langle p(\mathbf{x},r)\rangle_{\mathbf{x}\in \mathbb{R}^d}=\int_{C_{0,\dots, 0}} p(\mathbf{x},r) d\mathbf{x}
    \label{Ev_URL_alld}
\end{equation}
and
\begin{equation}
    p(\mathbf{x},r)=\prod_{|i_m|\leq \sqrt{d} + 1 ; m=1, \dots, d}|C_{i_1,\dots, i_d}\setminus\mathcal{B}(\mathbf{x};r)|
    \label{pqr_reduced_alld}
\end{equation}
respectively.

\section*{B Optimized effective potentials for the target models}
\label{append_opt}
For the 2D perfect glass, the $\alpha=1$ behavior of $S(k)$ at small $k$ implies that $v(r;\mathbf{a})\sim 1/r$ as $r\rightarrow\infty$. The fact that $g_2(r)=0$ for $r\leq 0.88$ implies that there is an effective hard core in this range. The intermediate-$r$ behavior of $v(r;\mathbf{a})$ was determined by fitting $v_{\text{HNC}}(r)$ in the range $0.88< r \leq 4$. No re-selection of basis function was needed. The optimized $v(r)$ is given by
\begin{equation}
v(r;\mathbf{a})=
\begin{cases}
\infty \qquad 0\leq r\leq 0.88\\
\frac{\varepsilon_1}{r}+\varepsilon_2\exp\left(-\frac{r}{\sigma_2^{(1)}}\right) \cos\left(\frac{r}{\sigma_2^{(2)}}+\theta_2\right)\\+\varepsilon_3 \exp\left(-(\frac{r}{\sigma_3})^8\right) \qquad r>0.88.
\end{cases}
\end{equation}
The optimized parameters are listed in Table \ref{table:pg}. The $L_2$-norm error is $\mathcal{E}=0.077$. 

For the 3D critical-absorbing-state model, the fact that $\alpha=0.25$ implies that $v(r;\mathbf{a})\sim 1/r^{2.75}$ at large $r$. Since the target system is a packing of spheres with unit diameter, $v(r;\mathbf{a})$ has a hard core for $r\leq 1$. The intermediate-$r$ behavior of $v(r;\mathbf{a})$ was obtained by fitting $v_{\text{HNC}}(r)$ in the range $1< r \leq 4$. No re-selection of basis function was needed. The optimized $v(r)$ is given by
\begin{equation}
v(r;\mathbf{a})=\begin{cases}
\infty \qquad 0\leq r\leq 1\\
\frac{\varepsilon_1}{r^{2.75}}+\varepsilon_2\exp\left(-(\frac{r}{\sigma_2})^5\right)+\varepsilon_3 \exp\left(-(\frac{r}{\sigma_3})^8\right) \\+ \varepsilon_4 \exp\left(-(\frac{r}{\sigma_4})^{10}\right) \qquad r>1.
\end{cases}
\end{equation}
The optimized parameters are listed in Table \ref{table:ro}. The $L_2$-norm error is $\mathcal{E}=0.066$.

The optimized potential for the 3D cloaked URL is given in Ref. \citenum{To22}.

\begin{center}
\begin{table}[htp]
\caption{Optimized parameters of the effective pair potential for the 2D perfect-glass model.}
\centering
\begin{tabular}{|c c|c c|} 
 \hline
 $\varepsilon_1$ & 4.572 & $\theta_2$ & 1.032 \\
 $\varepsilon_2$ & 12.26 & $\varepsilon_3$ & 5.803 \\
 $\sigma_2^{(1)}$ & 0.3921 & $\sigma_3$ & 0.9766 \\
 $\sigma_2^{(2)}$ & 0.1505 & & \\
 \hline
\end{tabular}
\label{table:pg}
\end{table}
\end{center}

\begin{center}
\begin{table}[htp]
\centering
\caption{Optimized parameters of the effective pair potential for the 3D critical-absorbing-state model.}
\begin{tabular}{|cc|cc|} 
 \hline
 $\varepsilon_1$ & 0.3400 & $\sigma_3$ & 1.582 \\
 $\varepsilon_2$ & -0.8120 & $\varepsilon_4$ & -49.13 \\
 $\sigma_2$ & 0.9949 & $\sigma_4$ & 0.8827 \\
 $\varepsilon_3$ & 0.1068 & & \\
 \hline
\end{tabular}
\label{table:ro}
\end{table}
\end{center}

\section*{C Inherent structures and ground states}
\label{append_gs}
\begin{table*}[htp]
    \centering
    \caption{Inherent-structure and ground-state energies levels for hyperuniform and nonhyperuniform models with $N=1000$ and the small-wavevector behaviors of the inherent structures.}
    \begin{tabular}{ccccccc}
    \hline
    System & Ground-state structure & $\overline{E}_0/\varepsilon$ & $\overline{E}_{\text{IS}}/\varepsilon$ & $\overline{E}_{\text{IS}}/\overline{E}_0$ & $S_{\text{IS}}(0)$ & $\alpha_{\text{IS}}$\\
    \hline 
    2D perfect glass & Fig. \ref{fig:gs}(a) & -5.973 \footnotemark[\value{footnote}] & $-5.594\pm 0.0033$ \footnotemark[\value{footnote}] & 94\% &0 & 2\\
    3D cloaked URL & sc & -1.081 \footnotemark[\value{footnote}]& -1.081 \footnotemark[\value{footnote}] & 100.0\% & 0 & $\infty$ \\
    3D critical-absorbing state & Fig. \ref{fig:gs}(b) & -7.104 \footnotemark[\value{footnote}]& $-5.29\pm0.035$ \footnotemark[\value{footnote}] & 74\% & 0 & 2\\
    3D GC & bcc \cite{Pr05} & 2.284297 & $2.284521\pm4.1\times 10^{-6}$ & 100.010\%& $8.9\times10^{-8}$ & \dots\\
    3D PL & fcc \cite{Tr14} & 6.066 & $6.971\pm0.026$ & 115\%& 0.0018 & \dots\\
    3D LJ & hcp \cite{Tr14} & -8.593 & $-7.836\pm0.018$ & 91\% & 0.0047 & \dots\\
    \hline
    \end{tabular}
    \label{tab:energies}
\end{table*}
Here, we present details on the energy levels of the inherent structures for the hyperuniform models considered in this study, including the 2D perfect glass, the 3D critical-absorbing state and the 3D cloaked URL. We also consider inherent structures for several well-studied 3D models that yield nonhyperuniform states at positive $T$ in three dimensions, including Lennard-Jones (LJ), inverse power-law (PL) and Gaussian-core (GC) \cite{St81} models. For all models, we ascertain the the inherent structures using the low-storage BFGS algorithm, \cite{Liu89} staring from equilibrium initial states away from phase transitions. The inherent-structure energies per particle $\overline{E}_{\text{IS}}$ are compared with the corresponding ground-state energies $\overline{E}_0$.

\footnotetext{Energy value includes the ``neutralizing'' one-body potential.}

\begin{figure}[htp]
  \centering
  \subfloat[]{\includegraphics[width=30mm]{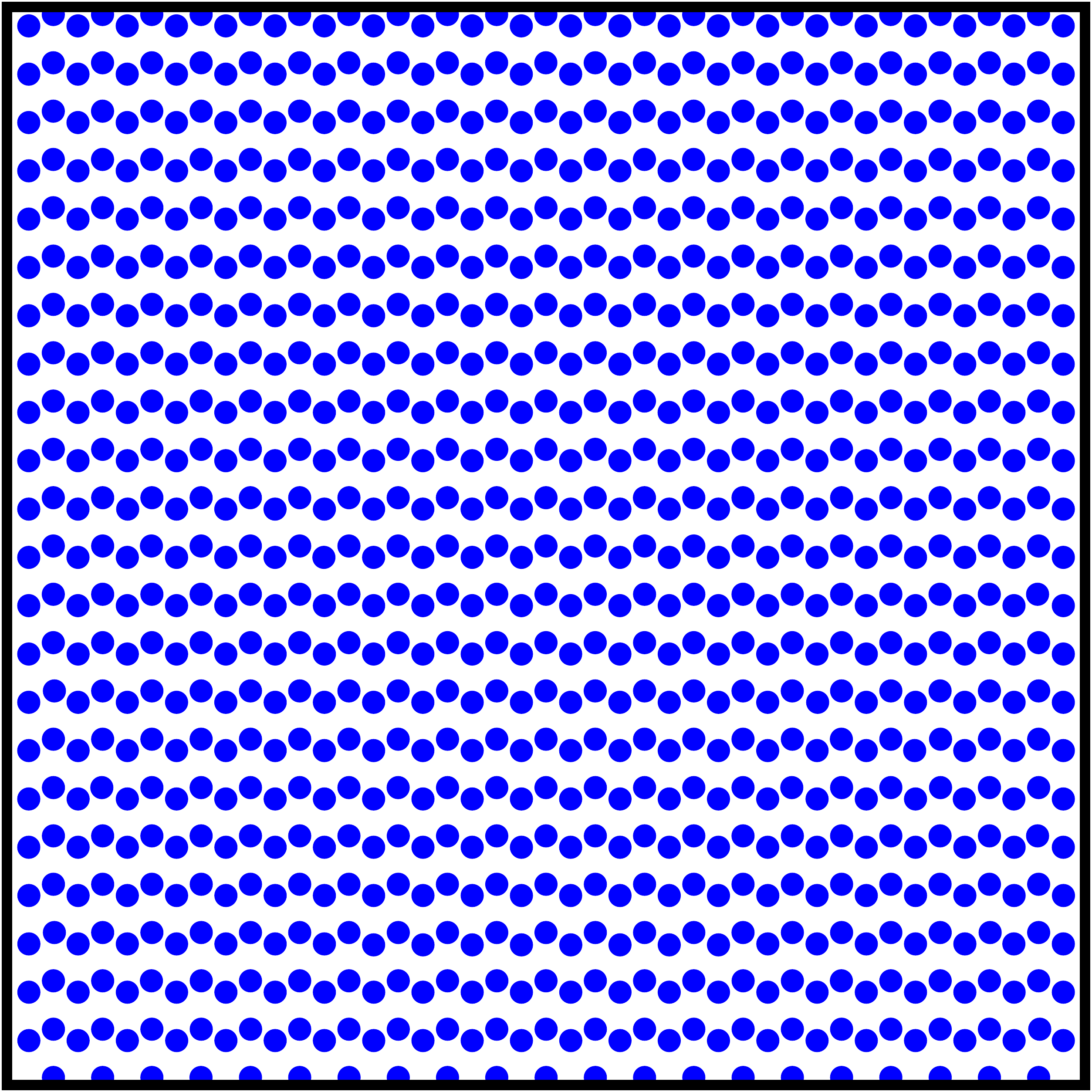}}
  \hspace{1mm}
  \subfloat[]{\includegraphics[width=40mm]{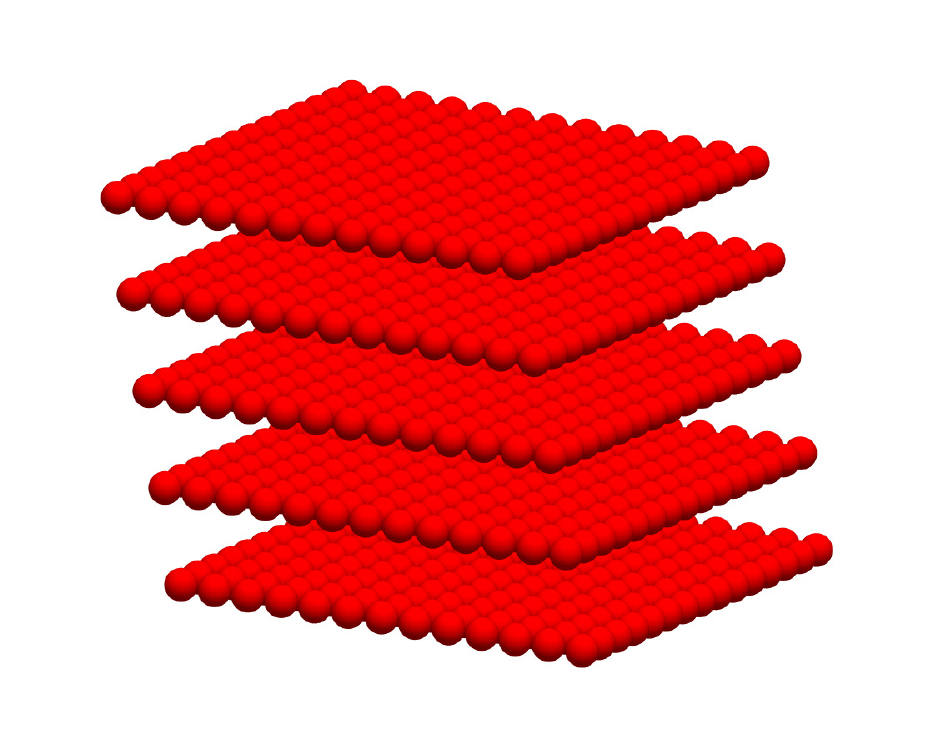}}
  \caption{Structures of the ground-state candidates associated with $v(r)$ for (a) the perfect glass and (b) the critical-absorbing state.}
  \label{fig:gs}
\end{figure}

To find the ground states for the hyperuniform models, we first applied the simulated annealing algorithm \cite{Ki83} on configurations with $N\geq 100$ particles, starting from the equilibrium liquid states that match the target pair statistics. The ground state for the URL effective potential is found to be the SC lattice. However, for the perfect glass and the critical-absorbing state, we found that even with very slow cooling rates, the annealing procedures resulted in disordered metastable states and were unable to find the global minima.\footnote{With cooling rate $T_{i+1}=0.999T_i$, where $i$ is the iteration index, the annealing procedures still result in metastable states for the hyperuniform models.} Therefore, we searched for ground-state candidates by optimizing \textit{crystalline} structures via simulated annealing over $n$-particle bases, where $n=1,2,\dots,10$. The parameters subject to optimization are unit cell vectors and coordinates of the $n-1$ particles in the interior of a unit cell. We observe that $v(r)$ for the perfect glass gives the same optimized crystal structure for all even $n$, whose energy is below the optimized energies for odd $n$, and that $v(r)$ for the critical-absorbing state yields identical optimized structures for all $n$. Thus, it is highly likely that the ground state for the perfect-glass potential is a two-particle basis, whereas that for the critical-absorbing-state potential is a Bravais lattice. Figure \ref{fig:gs} shows the ground-state candidates for both models. The ground-state candidate for the perfect glass contains zig-zag chains of particles, whereas the one for the critical-absorbing state contains planar sheets of triangle lattice.

The potentials for the nonhyperuniform models are given by
\begin{equation}
    v_{\text{LJ}}(r) = 4\varepsilon\left[(r/\sigma)^{-12}-(r/\sigma)^{-6}\right],
    \label{LJ}
\end{equation}
\begin{equation}
    v_{\text{PL}}(r) = 4\varepsilon\left[(r/\sigma)^{-12}\right],
    \label{PL}
\end{equation}
\begin{equation}
    v_{\text{GC}}(r) = \varepsilon\exp\left[-(r/\sigma)^2\right],
    \label{GC}
\end{equation}
where $\varepsilon$ and $\sigma$ are energy and distance scales, respectively. The LJ and PL models are strongly repulsive at small $r$. While $v_{\text{GC}}(r)$ is a soft interaction, the model possesses an effective hard core at low temperatures. \cite{St76, Za08} We compute the inherent structures for Eqs. (\ref{LJ})--(\ref{GC}) starting from equilibrium configurations at $\rho\sigma^3=1, kT/\varepsilon=1$. These $\rho,T$ values are chosen so that the initial states are dense liquids away from phase transitions for all three potentials. \cite{Tr14,Pr05,Za08}

Table \ref{tab:energies} lists values of $\overline{E}_0$ and $\overline{E}_{\text{IS}}$ for the hyperuniform and nonhyperuniform models with $N=1000$, as well as the small-wavevector behaviors extracted using the diffusion spreadability. \cite{Wa22a} For all models except the URL, $\overline{E}_{\text{IS}}$ values are narrowly distributed above $\overline{E}_0$. The distributions are Gaussian and are insensitive to the system size. For the perfect glass, $\overline{E}_{\text{IS}}/\overline{E}_0 = 94\%$,
and the standard deviation $\sigma_{\text{IS}}$ of $\overline{E}_{\text{IS}}$ is $5\times 10^{-4}\overline{E}_0$.
For the critical-absorbing state, $\overline{E}_{\text{IS}}/\overline{E}_0 = 74\%$ and $\sigma_{\text{IS}} = 0.005\overline{E}_0$. The inherent structure for the URL is the SC lattice, identical to the ground state. The fact that the structure is crystalline (periodic) means that it is stealthy hyperuniform and hence $\alpha\to\infty$. \cite{To18a}

Importantly, the three systems that are hyperuniform at positive $T$ yield hyperuniform inherent structures with increased values of $\alpha$ compare to those of the equilibrium states at positive $T$, which is consistent with the proposition that the quench reflects stronger forms of hyperuniformity, as reported in Sec. \ref{sec:res_is} of the main article. In particular, $\alpha$ for the URL potential dramatically increases from 2 at unit temperature to infinity upon quenching. On the other hand, the pair potentials for nonhyperuniform systems yield nonhyperuniform inherent structures. Note that while the inherent structure for the GC model is nearly hyperuniform, \cite{To18a} it is not perfectly hyperuniform, which again is consistent with the proposition stated in the main article.

%%%END OF MAIN TEXT%%%



%%%REFERENCES%%%
\bibliographystyle{unsrtnat}
%\bibliography{new} %You need to replace "rsc" on this line with the name of your .bib file

\end{document}